\documentclass[10pt]{article}
\usepackage{a4wide,epsfig,amsmath}

\begin{document}
\title{\bf ON THE INTEGRABILITY, B\"ACKLUND TRANSFORMATION AND SYMMETRY 
ASPECTS OF A GENERALIZED FISHER TYPE NONLINEAR REACTION-DIFFUSION EQUATION}
\author{P.S. BINDU$^{\dag}$, M. SENTHILVELAN$^{\ddag}$ and M. LAKSHMANAN$^{\dag}$\\
$^\dag$ Centre for Nonlinear Dynamics,
Department of Physics, Bharathidasan \\ University, Tiruchirapalli 620 024,
India\\
$^\ddag$ School of Physics, The University of Sydney, 
NSW 2006, Australia}
\date{}

\maketitle
\begin{abstract}

The dynamics of nonlinear reaction-diffusion systems is dominated  by the onset
of patterns and Fisher equation is considered to be a  prototype  of such
diffusive equations. Here we investigate the  integrability  properties of a
generalized Fisher equation in both  (1+1) and (2+1)  dimensions. A Painlev\'e 
singularity structure analysis singles out  a special  case ($m=2$) as
integrable. More  interestingly, a B\"acklund transformation is shown  to  give
rise to  a linearizing transformation for the integrable case. A  Lie symmetry 
analysis again separates out the same $m=2$ case as the  integrable one  and
hence we report several physically interesting  solutions via similarity
reductions. Thus we give a  group theoretical  interpretation for the system
under study. Explicit and  numerical solutions for specific cases of
nonintegrable  systems are also given. In  particular, the system is found to
exhibit different types  of travelling wave solutions and patterns, static 
structures and  localized structures. Besides the Lie symmetry  analysis, 
nonclassical  and generalized conditional symmetry analysis are also carried
out. 

\end{abstract}

\section{Introduction}
\label{sec1}
Integrable systems play the role of prototypical examples to identify and 
understand the various phenomena underlying nonlinear dispersive systems such 
as Korteweg-de Vries, sine-Gordon, Heisenberg spin, nonlinear Schr\"odinger,
Davey-Stewartson, Kadomtsev-Petviashivili, etc. equations 
[Ablowitz \& Clarkson, 1991; Lakshmanan \& Rajasekar, 2003]. 
Nonintegrable perturbations can be analysed in terms of the basic 
excitations
of such integrable models [Scott, 1999]. These integrable systems are often 
shown to
be linearizable in the sense that they can be associated with two linear
differential equations (of which one is a linear eigenvalue problem), namely the 
so called Lax pairs. In (1+1) dimensions, the exponentially localized 
stable 
entity, namely the soliton, turns out to be  the basic structure 
exhibiting 
elastic collision property. The soliton excitation has remarkable physical 
and 
mathematical properties. In particular, the underlying nonlinear evolution
equations exhibit infinite number of Lie-B\"acklund 
symmetries [Bluman \& Kumei, 1989]. 
The Lie
point symmetries lead to similarity reductions which are related to 
Painlev\'e 
type ordinary differential equations (ODEs) which are free from 
movable critical 
singular points. More generally, the integrable soliton equations
satisfy the Painlev\'e property and the solutions are in general
free from movable critical singular manifolds.

In the case of nonlinear reaction diffusion equations, the dynamics is
dominated by the onset of patterns [Murray, 1989; Walgraef, 1996; 
Lakshmanan \& Rajasekar, 2003]. Out of all 
possible 
modes available to the system, 
 they tend to select the most stable structures which 
give rise to various patterns. These patterns range from simple to chaotic,
depending on the nature of diffusion, nonlinear reaction terms and external 
forces as well as the spatial dimension. Some of the dominant patterns 
are \\
(1) homogeneous or uniform states \\ (2) travelling waves \\
(3) wavefronts and pulses \\ (4) Turing structures: a) stripes b) spirals
c) scrolls, etc. \\ (5) spatiotemporal chaos \\
and so on. Thus mode selection and stability dominates the study of such 
systems. Some of the well known nonlinear diffusion and 
reaction-diffusion systems[Murray, 1989; Walgraef, 1996; Whitham, 1974] include 
Burgers' equation, 
Fisher equation, 
Kuromoto-Sivashinsky equation, Gierer-Meinhardt equation, FitzHugh-Nagumo
equation, Belousov-Zhabotinsky reaction equation, Brusselator model equation 
and so on. A large amount of literature on such systems is available.

In view of the complexity involved in analysing nonlinear diffusion equations, 
it will be very valuable to identify integrable nonlinear 
diffusive systems and to build on them 
the study of nonintegrable cases. Burgers' equation is a standing example of 
such an integrable case [Sachdev, 1987]. It is linearizable in the sense that 
the Burgers' equation
\begin{eqnarray}
\label{int1}
u_t=Du_{xx}-uu_x,
\end{eqnarray}
under the Cole-Hopf transformation
\begin{eqnarray}
\label{int2}
u=-2D\frac{v_x}{v}
\end{eqnarray}
gets transformed into the linear heat equation
\begin{eqnarray}
\label{int3}
v_t=Dv_{xx}.
\end{eqnarray}
Similarly the Fokas-Yortsos-Rosen equation [Fokas \& Yortsos; 1982 \& Rosen, 1982]
\begin{eqnarray}
\label{int4}
u_t=u^2(Du_{xx}-u_x)
\end{eqnarray}
under the variable transformations [Rosen, 1982] 
\begin{eqnarray}
\label{int5}
x=- D\ln v, u= -D \frac{v_x}{v},
\end{eqnarray}
gets again transformed into the linear heat equation 
Eq.~(\ref{int3}). Thus both the above 
systems are linearizable and they may be considered to be 
C-integrable in the Calogero [1991]
sense. Also both the Burgers' equation and Fokas-Yortsos-Rosen 
equation
possess interesting Lie point symmetry structures and infinite 
number of 
Lie-B\"acklund symmetries. So, it will be of interest to 
investigate whether
other integrable and linearizable nonlinear reaction-diffusion equations exist 
and if so what is the
role of symmetry  therein and what kind of solution structures 
and patterns exist in them.

Considering nonlinear reaction-diffusion equations, Fisher 
equation is 
considered to be prototypical [Murray, 1989]. Typically, its form 
reads as
\begin{eqnarray}
\label{int6}
u_t - \triangle u  - u + u^2 = 0,
\end{eqnarray}
where $\triangle$ is the Laplacian operator.
In one dimension it admits travelling wavefronts with velocities 
$c\ge c_{min} = 2$, depending on the initial 
condition [Murray, 1989] and in two 
dimensions interesting spatiotemporal patterns arise; however
\emph{it is 
not integrable in any dimensions}. In one dimension, however it 
is known 
to have an exact propagating wavefront solution of the 
form [Ablowitz \& Zeppetella, 1979]
\begin{eqnarray}
\label{int7}
u(x,t)=1-\Bigg[1+\frac{k}{\sqrt{6}}
\exp\Bigg(\frac{x-\frac{5}{\sqrt6}t}{\sqrt 6}\Bigg)\Bigg]^{-2},
\end{eqnarray}
where $k$ is an arbitrary constant, corresponding to the wave speed 
$c=\sqrt{5}/6$. No other exact solutions are known for Eq.~(\ref{int6}). 

\section{Generalized Fisher (GF) Equation}
\label{sec2}
In this connection, it has been pointed out [Grimson \& Barker, 1994] that there 
arises interesting 
generalization of the Fisher equation in the description of bacterial  
growth in multicolony environment, in chemical kinetics and in various living
phenomena. Its general form is
\begin{eqnarray}
\frac{\partial u({\bf r},t)}{\partial t} = D\triangle 
u({\bf r},t)
+\Lambda(u)[\nabla u({\bf r},t)]^2+\lambda u({\bf r},t)
G(u),
\label{gfe1}
\end{eqnarray}
where $D$ is the diffusion coefficient, 
$\nabla$ and $\triangle$ are 
gradient and Laplacian operators respectively. To be specific,
Eq.~(\ref{gfe1}) 
represents the spatiotemporal growth of bacterial colonies 
with both local and nonlocal modes of growth. The third term on 
the right-hand side of Eq.~(\ref{gfe1}) represents the local
growth characterised by the local growth rate $\lambda$ and a 
local growth function $G(u)$. Suppose the colony
is unable to add cells locally due 
to packing constraints, 
the nonlocal addition of cells on the surface of the colony
occurs to account for the amount of growth required by the global
growth law. Such an addition occurs as a consequence of the 
expansion and reorganization of the colony and is represented
by the second term while the first term represents diffusion.

An important special case of Eq.~(\ref{gfe1}) is the generalized Fisher 
type equation 
\begin{eqnarray}
\label{gfe2}
u_t - \triangle u-\frac{m}{1-u} ({\nabla u})^2 - u(1-u)=0,
\end{eqnarray}
where the subscript denotes partial differentiation with respect to time.
In the study of population dynamics, $u({\bf r},t)$ refers to the 
population density at the point ${\bf r}$ at time $t$. 
In Eq.~(\ref{gfe2}), the nondiffusive linear 
term modelling the birth rate gives rise to an exponential 
growth of $u({\bf r},t)$ in time while
the quadratic term that models competition between individuals for 
food, etc. 
leads to a stable, homogeneous value $u=1$ at long times and the 
diffusion term
represents the spatial variation of the population. 
Further  the third term represents 
the nonlocal growth of cells while the fourth
term corresponds to the local growth. This introduces the possibility 
of spatial pattern formation between the homogeneous regions with $u=1$ and 
$u=0$ for appropriate initial conditions. Note that for $m=0$, 
Eq.~(\ref{gfe2}) reduces to the standard Fisher equation (\ref{int6}). Again
Eq.~(\ref{gfe2}) is in general nonintegrable, except for the special case 
$m=2$. It is linearizable under the transformation 
\begin{eqnarray}
u=1-\frac{1}{1+\chi},
\label{gfe3}
\end{eqnarray} 
as pointed out by Wang \emph{et al.} [1996] for the one dimensional
case (see also Brazhnik and 
Tyson [1999a,b]). Under this transformation the generalized Fisher 
equation (\ref{gfe2}) can be shown to reduce to the form
\begin{eqnarray}
\chi_t-\triangle \chi-\chi=0,
\label{gfe4}
\end{eqnarray}
which is nothing but the linear heat equation in arbitrary dimensions. Thus
every solution of (\ref{gfe4}) corresponds to a solution of Eq.~(\ref{gfe2}) 
via the transformation (\ref{gfe3}) subject to boundary and initial 
conditions. Note that the transformation holds good irrespective of the 
spatial dimensionality of the generalized Fisher system 
(\ref{gfe2}) as long 
as the parameter takes the value $m=2$. Such a 
\emph{linearization 
in any dimensions is a rare situation} indeed. Even in soliton theory,
straightforward extensions of integrable equations lose their integrability
as soon as the dimension is increased. In fact Wang \emph{et al.} [1996] and 
Brazhnik and  Tyson [1999a,b] make use of this linearizability property
to obtain several interesting wave solutions, spatiotemporal patterns and 
static structures.

The very special nature of the $m=2$ case prompts us to investigate the
integrability nature of the generalized Fisher equation (\ref{gfe2}) from  
different points of view, particularly from singularity structure
and symmetry points of 
view in order to understand why the case $m=2$ alone is 
integrable. For the nonintegrable case ($m\ne2$) we investigate 
special solutions.

In Sec. \ref{sec3}, we show that the PDE (\ref{gfe2}) in one 
dimension is free from movable critical singular manifolds only for the 
special case $m=2$. Further we point out that a B\"acklund 
transformation gives rise to the linearizing transformation in a 
natural way for the integrable case. In Sec. \ref{s4}, we derive 
certain explicit and numerical solutions for the GF equation in 
one dimension via symmetry analysis and similarity reductions. In 
Sec. \ref{s5}, we present the associated Lie algebra for the (2+1) 
dimensional GF equation. Then we report different types of 
propagating structures, static solutions and localized structures
through Lie symmetry analysis in Secs. \ref{s6} and \ref{s7}.  
In addition, exact solutions 
for special cases of the nonintegrable GF equation are also given.
We give the nonclassical and generalized conditional
reductions in Secs. \ref{s8} and \ref{s9} respectively.  Finally, we summarise
our results in  Sec. \ref{con}.

\section{Singularity  Structure Property of the GF Equation}
\label{sec3}
It has been argued for sometime that the Painlev\'e property 
(P-property) as 
proposed by 
Weiss \emph{et al.} [1983] is a working test to
identify whether  a given nonlinear partial differential equation (PDE) is 
integrable or not. The Painlev\'e property  demands that the solution of the
given nonlinear PDE is free from movable critical singular manifolds 
(branching types, both algebraic and logarithmic as well as 
essential singular),
so that the solution is single valued in the neighbourhood of a  
noncharacteristic movable singular manifold, $\phi({\bf r},t)= 0$, 
$\phi_x,\phi_y,\phi_t \ne 0$. The Weiss Tabor Carnevale (WTC) procedure then demands that the 
solution $u({\bf r},t)$ for Eq.~(\ref{gfe2}) can be expanded 
around the noncharacteristic movable singular manifold as a Laurent 
expansion, 
\begin{eqnarray}
\label{pain1}
u=\sum_{j=0}^\infty u_j({\bf r},t) \phi^{j+p}
\end{eqnarray}
where $u_j$'s are functions of ${\bf r}$ and $t$ 
and $p$ is the leading order exponent of the Laurent expansion. 
For single 
valuedness of the solution, one requires that $p$ is an integer 
and that
there exist sufficient number of arbitrary functions, 
$u_j({\bf r},t), j=1,2,...,N$ (for Eq.~(\ref{gfe2}) it is two), so
that the Laurent expansion (\ref{pain1}) can be considered as the
general solution of the nonlinear PDE (\ref{gfe2}) without the 
introduction of  
movable critical singular manifolds. Following the standard 
algorithm we can proceed as follows. As an example, we demonstrate 
the analysis
for the one-dimensional case, which can be then extended to the 
higher 
dimensional case without difficulty.
\subsection{\bfseries \emph{P-property of eq.~(\ref{gfe2})}}
\label{subsec2a}
In the one-dimensional case, the generalized Fisher equation can 
be written as
\begin{eqnarray}
u_t-u_{xx}-\frac{m}{1-u}u_x^2-u+u^2=0.
\label{pain2}
\end{eqnarray}
The WTC algorithm then identifies the following results.
Starting with the leading order behaviour $u\approx u_0\phi^p$ 
as suggested by the form (\ref{pain1}), one can identify  three 
possibilities:
\begin{eqnarray}
 (i) \mskip 10mu  p &=& -2 \mskip 400mu \nonumber \\
(ii) \mskip 10 mu p &=& \frac{1}{1-m},\mskip 10mu m \ne 1 \mskip 
400mu\nonumber
  \\
(iii) \mskip 10 mu  p &=& 0.\mskip 400mu
\label{pain3}
\end{eqnarray}
One can easily check that for all the above three leading orders, only for
the value $m=2$, the solution is free from movable critical singular manifolds.
Actually one finds that for $m=2$ and  $p=-1$, the leading order coefficient 
$u_0$ is an arbitrary function in addition to the arbitrary nature of the 
singular manifold so that the Laurent series contains the required number of 
arbitrary functions in terms of which all the other coefficient functions 
$u_i(x,t), i\ge 1$, can be uniquely found. For the other two leading orders 
with $m=2$, one identifies only one arbitrary function but without the 
introduction of any movable critical singular manifolds and so they may be 
considered as  corresponding to special solutions. The above analysis also 
holds good in higher dimensions of (\ref{gfe2}) as well.

Consequently one finds that the P-property is satisfied only for the special 
choice of parameter, $m=2$ for Eq.~(\ref{gfe2}) and for all other values of 
$m(\ne2)$ it fails to satisfy the P-property and the system is expected to be 
nonintegrable. Note however that there may exist certain special 
singular manifolds $\phi_t-\phi_{xx}=0$, for which the P-property may be 
satisfied in special situations. However in general for the special choice of
the parameter $m=2$ of Eq.~(\ref{gfe2}) the P-property is 
satisfied for general singular manifolds and 
the system is expected to be integrable in this case.

\subsection{\bfseries\emph {B\"acklund transformation}}
\label{ssecp}
Now we consider the Laurent series (\ref{pain1}) applicable to 
Eq.~(\ref{pain2})
and truncate it at constant level term, by making all other 
coefficient functions $u_i({\bf r},t)$ to vanish for $i\ge 1$ consistently:
\begin{eqnarray}
u=\frac{u_0}{\phi}+u_1.
\label{pain4}
\end{eqnarray}
We demand that if $u_1$ is a solution of (\ref{pain2}), so also 
$u$ is a solution. Then, from
Eq.~(\ref{pain4}), we have 
\begin{eqnarray}
 u_{1t}-u_{1xx}-u_1(1-u_1)+(u_{0t}-u_{0xx}-u_0(1-2u_1))\phi^{-1}+
(2u_{0x}\phi_{x}\nonumber &\\-u_0(\phi_t-\phi_{xx}-u_0)\phi^{-2}
-2u_0\phi_x^2\phi^{-3}-2\phi{\bigg[(1-u_1)\phi-u_0
\bigg]}^{-1}\Big[u_{1x}^2\Big .\nonumber &\\ \Big .+2u_{0x}u_{1x}\phi^{-1}
+(u_{0x}^2-2u_0u_{1x}\phi_x)\phi^{-2}-2u_0u_{0x}\phi_x
\phi^{-3}+u_0^2\phi_x^2\phi^{-4}\Big]&=0.
\label{pain5}
\end{eqnarray}
 
Consequently one finds that $u_0$ and $\phi$ satisfy a set of coupled PDEs 
arising from Eq.~(\ref{pain5}). Starting from the trivial solution, $u_1=0$,
of Eq.~(\ref{pain2}), we find that the choice 
\begin{eqnarray}
u_0=\phi
\label{pain6}
\end{eqnarray}
solves Eq.~(\ref{pain5}). As a result, we find the new solution
\begin{eqnarray}
u=\frac{u_0}{\phi}+u_1=1
\label{pain7}
\end{eqnarray}
which is indeed a solution to Eq.~(\ref{pain2}).

Taking the solution (\ref{pain7}) as $u_1$ now, we find that the system 
(\ref{pain5}) satisfies  the equations
\begin{eqnarray}
u_0&=&-1, \nonumber\\
\phi_t-\phi_{xx} -\phi +1 &=&0. 
\label{pain8}
\end{eqnarray}
Now defining the new function $\chi$ as 
$\phi=1+\chi$ one obtains the linear heat equation
\begin{eqnarray}
\chi_t- \chi_{xx} - \chi=0.
\label{pain9}
\end{eqnarray}
Thus the transformation [Bindu \emph{ et al.}, 2001; Bindu \& Lakshmanan, 2002] 
\begin{eqnarray}
u=1-\frac{1}{1+\chi},
\label{pain10}
\end{eqnarray}
is the \emph{linearizing transformation} for Eq.~(\ref{pain2}), where 
the function
$\chi$ satisfies the linear heat equation (\ref{pain9}).
We note that this is exactly the
transformation given by Wang \emph{et al.} [1996] in an adhoc way. Here
the transformation is given an   
 interpretation in terms of the \emph{B\"acklund 
transformation}. Thus the application of Painlev\'e singularity structure 
analysis selects not only the integrable case associated with
 Eq.~(\ref{pain2}), 
but also helps to identify the associated linearizing transformation.

One can generalize the above analysis to higher dimensional case of the Eq.~(\ref{gfe2}) for the choice $m=2$ and arrive at the same general conclusion that the transformation (\ref{pain10}) can be 
interpreted as a B\"acklund transformation.
\section{Lie Point Symmetries of the (1+1) Dimensional GF Equation and 
Integrability}
\label{s4}
In this Sec., we investigate group invariance properties of the 
generalized Fisher equation in its (1+1) dimensional form given in 
Eq.~(\ref{pain2}) 
and discuss different underlying patterns via similarity reductions.  First 
we consider the invariance of Eq.~(\ref{pain2}) under the one-parameter 
Lie group of infinitesimal transformations and derive the similarity 
reduced ODEs for integrable ($m=2$) and nonintegrable ($m\neq 2$) cases 
separately.  We solve the ODEs and provide explicit solutions.  We  
show that the earlier reported solutions are sub-cases of our results.  We 
construct the linearizing transformation from the infinitesimal symmetries 
and provide a group theoretical interpretation for it. 
\subsection{\bfseries \emph{Lie symmetries}}
\label{sec4a}
An invariance analysis of Eq.~(\ref{pain2}) under the 
one-parameter Lie group of 
infinitesimal transformations

\begin{eqnarray}
&t \longrightarrow T = t+\varepsilon\tau(t,x,u), \quad x \longrightarrow  X = x+\varepsilon\xi(t,x,u),\nonumber&\\
&u \longrightarrow  U = u+\varepsilon\phi(t,x,u),\quad \varepsilon \ll1,&
\label{ls1}
\end{eqnarray}
singles out a special value of $m$, namely
$m=2$, to have  nontrivial infinite-dimensional
Lie algebra of symmetries
\begin{equation}
 \tau  =  a, \quad \xi  =   b, \quad \phi  = c(t,x)(1-u)^2,
\label{ls2}
\end{equation} 
where $ a, b $ are arbitrary constants and  $ c(t,x) $
is any solution of the linear heat equation $c_t-c_{xx}-c =0$.
The infinitesimal symmetries (\ref{ls2}) are obtained by following the 
usual procedure of solving the determining equations for the 
infinitesimal symmetries [Bluman \& Kumei, 1989]. Throughout this paper we 
use the computer program MUMATH [Head, 1993] to determine
infinitesimal symmetries and symmetry algebra.

The Lie algebra of infinitesimal symmetries 
of the Fisher equation is spanned by two commuting vector fields
and a vector field associated with an infinite-dimensional 
subalgebra,
\begin{equation} 
X_1 = \partial_t,\;\;
X_2 = \partial_x,\;\;
X_c = c(t,x){(1-u)}{^2} \partial_u \label{ls3},
\end{equation}
respectively.  In the above, the vector fields $X_1$ and $X_2$ reflect the 
time and spatial invariance of Eq.~(\ref{pain2}).  
The commutation relations between these vector fields are given by
\begin{eqnarray}
[X_1,X_2]=0,\quad [X_1,X_c]=X_{c_t},\quad  [X_2,X_c]=X_{c_x},
\label{ls4}
\end{eqnarray}
where $X_{c_t}=c_t(1-u)^2\partial_u$ and 
$X_{c_x}=c_x(1-u)^2\partial_u$.
For all other values of the parameter $m(\ne2)$, Eq.~(\ref{pain2}) admits 
only the trivial translational symmetries
\begin{equation}
\tau = a,\quad \xi = b,\quad \phi = 0,
\label{ls5}
\end{equation} 
with the corresponding symmetry algebra
\begin{eqnarray}
[X_1,X_2]=0,
\label{ls6}
\end{eqnarray}
where $X_1=\partial_t$ and  $X_2=\partial_x$.
\subsection{\bfseries \emph{ Similarity reductions}}
\label{ssecb}
 We solve the characteristic 
equation associated with the symmetries (\ref{ls2}) and 
(\ref{ls5}) and obtain similarity variables in terms of which 
Eq.~(\ref{pain2}) can be reduced to an ODE. From the general 
solution of the ODE we construct explicit solutions for the 
GF equation (\ref{pain2}).

\subsubsection*{ 4.2(A)  \bfseries \emph{ Integrable case $(m=2)$}}
Solving the characteristic equation
\begin{eqnarray} 
{\frac {dt} {a} \quad = \quad \frac {dx}{b} \quad = \quad
\frac{du} {c(t,x)(1-u)^2}}
\label{ls7}
\end{eqnarray} 
associated with the symmetries (\ref{ls2}), we obtain the 
similarity reduced variables 
\begin{eqnarray} 
z=ax-bt,\quad  
u=1- \displaystyle {\frac {a}{a+v(z)+\int c(t,x)dt}}. 
\label{ls8}
\end{eqnarray}
In Eq.~(\ref{ls8}) we can consider the quantity
\begin{eqnarray}
\displaystyle{\frac{1}{a}}\Big(v(z)+\int c(t,x)dt\Big)=\chi,
\label{ls9}
\end{eqnarray}
and verify that $\chi$ satisfies the linear heat equation 
(\ref{pain9}). 
Then  the \emph{similarity transformation} given by 
Eq.~(\ref{ls8}) is 
the \emph{linearizing transformation} thereby 
giving a group theoretical interpretation of it.

Making use of (\ref{ls8}), Eq.~(\ref{pain2})  reduces
to  an ODE
\begin{eqnarray} 
\displaystyle{a^2 v''+bv' + v=0},\quad '=\frac{d}{dz},
\label{ls10}
\end{eqnarray}
whose general solution can be written as 
\begin{eqnarray}
v = I_1 e^{\displaystyle {{m_1z}}}+ I_2 e^{\displaystyle {{m_2z}}}, \quad 
m_{1,2}=\frac{-b \pm \sqrt {b^2-4a^2}}{2a^2}.
\label{ls11}
\end{eqnarray}
Here $I_1$ and $I_2$ are integration constants. Now with the use of 
Eq.~(\ref{ls8}), the solution to the original PDE (\ref{pain2}) becomes
\begin{eqnarray}
u= \left\{
	\begin{array}{lll}
1-\displaystyle{\frac {a}
{a+I_1 e^{\displaystyle{{m_1(ax-bt)}}}+I_2 e^{\displaystyle{{m_2(ax-bt)}}}+
\int c(t,x) dt}}, \;\;\qquad b^2-4a^2 > 0;
\\
1-\displaystyle{\frac{a}{ a+e^{\displaystyle{{p(ax-bt)}}} 
\left( I_1+I_2(ax-bt) \right)
+\int c(t,x)dt }},\;\; \;\;\;\;\;\;\qquad b^2-4a^2 = 0; \\
1-\displaystyle{\frac{a}{a+e^{\displaystyle{{p(ax-bt)}}} 
\left( I_1 \cos {q(ax-bt)} 
+ I_2 \sin {q(ax-bt)} \right)
+\int c(t,x)dt}},\\ \qquad  \mskip 440 mu b^2-4a^2 < 0,
\label{ls12}
\end{array}
\right. 
\end{eqnarray}
with $p = -b/2a^2$,$\;$ $q = \sqrt{4a^2-b^2}/2a^2$. 

We note the following: By fixing the
integration constants $I_1$ and $I_2$ suitably and for the 
choice 
$c(t,x)=0$  we can deduce all the interesting travelling wave 
solutions and stationary structures discussed by Brazhnik and 
Tyson [1999a,b]. For example, let us consider 
$b^2-4a^2>0$. 
Taking either one of the constants  $I_1$ or $I_2$ 
equal to zero we have a simple
travelling wave solution. The choice $I_1=I_2\ne0$ leads to 
a V-wave pattern. A Y-wave solution is produced by $I_1=-I_2$. An oscillating wave 
front is obtained by restricting   $I_2=0$ with $b^2-4a^2<0$. 
The choice $b^2-4a^2=0$ leads to  an inhomogeneous solution when $I_1=0, I_2\ne 0$.

\subsubsection*{ 4.2(B) \bfseries \emph{Nonintegrable case $(m \ne 2)$}}
A similar analysis on Eq.~(\ref{ls5}) for the $(m\ne2)$ case 
leads to the similarity variables
\begin{eqnarray}
z=ax-bt,\quad u=w(z),
\label{ls13} 
\end{eqnarray}
which reduces the original PDE (\ref{pain2}) to an ODE 
of the form
\begin{eqnarray} 
a^2vv'' - ma^2v'^2 + bvv' - v^2(1- v ) =0,\quad v  = 1-w.
\label{ls14}
\end{eqnarray} 
For $b\ne0$, this equation is in general nonintegrable except for $m=0$ and 
$b/a=5/\sqrt{6}$. This special choice leads to the cline solution
\begin{eqnarray}
\label{ls15}
u(x,t)=1-\Bigg[1+\frac{k}{\sqrt{6}}
\exp\Bigg(\frac{x-\frac{5}{\sqrt6}t}{a\sqrt 6}\Bigg)\Bigg]^{-2},
\end{eqnarray}
where $k$ is an arbitrary constant reported by Ablowitz and 
Zeppetella [1979] (Fig.~1) and the surface plot is shown in Fig.~2.
\begin{figure}[!ht]
\begin{center}
\epsfig{file=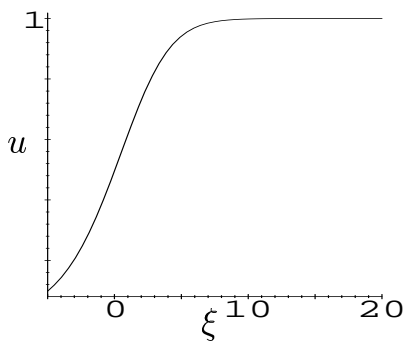, width = .35\linewidth}
\end{center}
\caption{ An exact wavefront solution of the Fisher equation   
($\xi = {x-\frac{5}{\sqrt6}t}$).}
\label{front}
\end{figure}

\begin{figure}[!ht]
\begin{center}
\epsfig{file=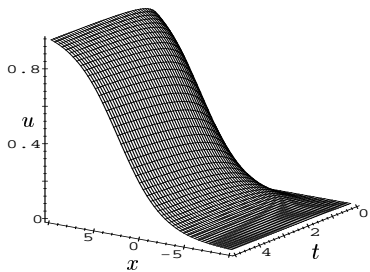,width=.35\linewidth}
\end{center}
\caption{Surface plot of the solution (\ref{ls15}).}
  \label{splot}
\end{figure}

{{\bfseries Case (a): Static solutions}} $(b=0,\;m\ne0)$

This is the time-independent (static) case and the similarity variables 
$z=x,\; w=u$ lead to the reduced ODE of the form
\begin{eqnarray}
w''+\frac{m}{1-w}w'^{2}+w(1-w) = 0.
\label{ls16}
\end{eqnarray}
This readily gives the first integral [Mathews \& Lakshmanan, 1974]
\begin{eqnarray}
v'^{2}=I_1v ^{2m}+\frac{1}{1-m}v ^2-\frac{2}{3-2m}v ^3, 
\quad v  = 1-w,\quad m\ne1,3/2,
\label{ls17}
\end{eqnarray}
 where $ I_1 $ is the integration 
constant. Eq.~(\ref{ls17})  on integration leads to elliptic
or hyperelliptic function solutions for suitable values of $m$.
For example, let us consider $m = 1/2$. 
Then Eq.~(\ref{ls17}) becomes
\begin{eqnarray}
v'^{2}=\left[I_1+2v -v ^2\right]v. 
\label{ls18}
\end{eqnarray}
Integration of Eq.~(\ref{ls18}) leads to
\begin{eqnarray}
\int \frac{dv }{\sqrt{v(I_1+2v-v^2)}} = z-I_2,
\label{ls19}
\end{eqnarray}
where $I_2$ is the  second integration constant.
Let $c_1 = 1+\sqrt{1+I_1}$, $\;$ $ c_2 = 1-\sqrt{1+I_1}$. Then 
Eq.~(\ref{ls19}) becomes
\begin{eqnarray}
\int \frac{dv }{\sqrt{(c_1-v)(c_2-v)(-v)}} = z-I_2.
\label{ls20}
\end{eqnarray} 
Now the LHS of the above equation can be integrated in terms of
 elliptic 
functions and the solutions are tabulated in Table~1.
In a similar way one can integrate Eq.~(\ref{ls17}) 
 and obtain elliptic and hyperelliptic function solutions for certain values of $m$.    
Besides the elliptic function solutions, for a particular value 
$I_1=0$ in Eq.~(\ref{ls18}), we obtain a static solitary wave solution 
(Fig.~3)
\begin{eqnarray}
u= 1-\displaystyle{\frac{(3-2m)}{(2-2m)}}\left[ \mbox{sech} ^2 
\left(I_2-\frac{x}{2}
\sqrt{\frac{1}{1-m}} \right) \right], \quad m < 1. 
\label{ls21}
\end{eqnarray}
Eq.~(\ref{ls21}) is nothing but a limiting case of the elliptic 
function solutions. 
\begin{figure}[!ht]
\begin{center}
\epsfig{file=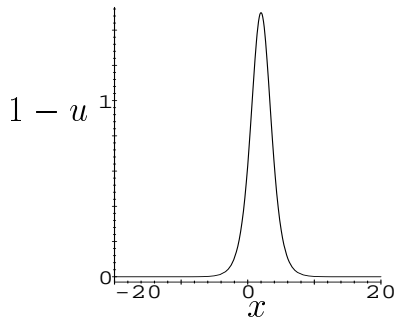, width=.35\linewidth}
\end{center}
\caption{ A static solitary wave pulse for $m=1/2$ of the generalized Fisher equation, 
Eq.~(\ref{ls21}).}
\label{soliton}
\end{figure}

{\bfseries Case (b): Qualitative analysis}

In general Eq.~(\ref{ls14}) is of nonintegrable type and 
this leads to
the use of numerical techniques to explore the 
underlying dynamics. Here we determine the 
equilibrium points and then study the system dynamics in the vicinity of these 
equilibrium points. 
Eq.~(\ref{ls14}), after a rescaling, can be written as 
\begin{eqnarray}
v'&=&p\nonumber\\
p'&=&m\frac{p^2}{v}-Bp+v(1-v), \; '=d/d\bar{z},
\label{qa1}
\end{eqnarray}
where $\bar{z}=z/a$ and $B=b/a$. Then the equilibrium points are found to be $(0,0)$ and $(1,0)$. While the origin is an unstable equilibrium point, the  stability determining eigenvalues 
for the equilibrium point (1,0) are found to be 
\begin{eqnarray}
\lambda_{1,2}=\frac{-B\pm\sqrt{B^2-4}}{2}.
\label{qa2}
\end{eqnarray}
Now the nature of the singular point $(1,0)$ is investigated as a function of the control parameter $B$ in the range $(-\infty,\infty)$ by analysing the form of the eigenvalues $\lambda_{1,2}$ given by 
Eq.~(\ref{qa2}) and the results are tabulated in Table~2 and the corresponding plots 
are shown in Fig.~(4). 
One identifies the existence of periodic pulses and wave front solutions in this case.
\begin{figure}[!ht]
\begin{center}
\epsfig{file=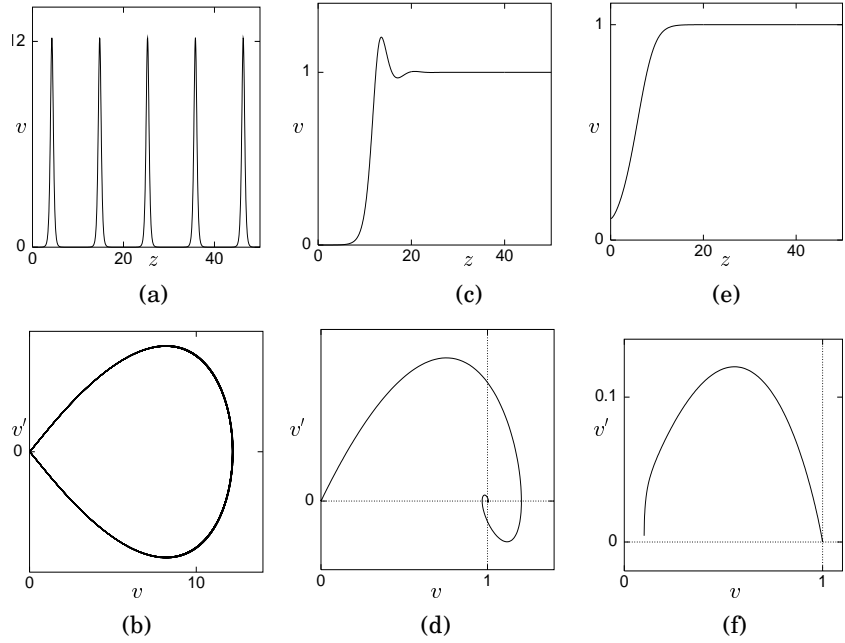, width=.75\linewidth}
\end{center}
\caption{ Propagating patterns and corresponding phase portraits in the 
$v-v'$ plane of Eq.~(\ref{ls14}) (a) Periodic pulses, 
(b) Center, (c) Travelling pulse, (d) Stable focus,
(e) Travelling wavefront, (f) Stable node. }
\label{phase}
\end{figure}  
\begin{table}[!ht]
\begin{center}
\caption{ Table 2. Classification of equilibrium point $(1,0)
$ of Eq.~(\ref{qa1}) as a function of the parameter $B$}
\end{center}
\begin{center}
\begin{tabular}{|c|c|c|c|}
\hline
S.No. & {Range of $B$} & Nature of eigenvalues & Type of attractor/repeller
 \\ \hline

1. & $-\infty<B<-2 $    & $\lambda_{1,2}>0,\; \lambda_1\ne\lambda_2$ & unstable node \\

2. & $B=-2 $    & $\lambda_{1,2}>0,\; \lambda_1=\lambda_2$ & unstable star \\

3. & $-2<B<0 $    & $\lambda_{1,2}=\alpha\pm i\beta,\; \alpha>0$ & unstable focus \\

4. & $B=0 $    & $\lambda_{1,2}=\pm i$, & center\\

5. & $0<B<2 $    & $\lambda_{1,2}=\alpha\pm i\beta,\;\alpha<0$ & stable focus \\
 
6. & $B=2 $    & $\lambda_{1,2}<0,\; \lambda_1=\lambda_2$ & stable star \\

7. & $2<B<\infty $    & $\lambda_{1,2}<0,\; \lambda_1\ne\lambda_2$ & stable node \\
\hline
\end{tabular}
\label{table2}
\end{center}
\end{table}

\subsection{ \bfseries \emph{Relation between symmetries of $m=2$  case and of linear heat 
equation}}
\label{sec4c}
In subsection \ref{ssecp}, we derived  a B\"acklund 
transformation that transforms the GF equation to a linear heat
equation via singularity structure analysis. In this Sec., by employing 
the ideas given by Bluman and Kumei, [1989] (Theorem 6.4.1-1,2, p.320), we derive the 
linearizing transformation from the symmetries and thereby confirming the group theoretical interpretation for
 it.  The 
aforementioned theorem 
gives a necessary and sufficient condition for the existence of an invertible 
mapping which linearizes a system admitting an infinite-parameter Lie group 
of transformation. That is, if a given nonlinear system of PDEs admit an infinite-parameter Lie group of transformation having an infinitesimal operator
\begin{eqnarray}
\label{lin1}
\hat{X}=\xi_i(x,u)\partial_{x_i}+\eta^\nu(x,u)\partial_{u^\nu},
\end{eqnarray}
with 
\begin{subequations}
\label{lin2}
\begin{eqnarray}
\xi_i(x,u)&=&\sum_{\sigma=1}^{m}\alpha_i^\sigma(x,u)F^\sigma(x,u),\\ \label{lin2a}
\eta^\nu(x,u)&=&\sum_{\sigma=1}^{m}\beta_\nu^\sigma(x,u)F^\sigma(x,u),
\label{lin2b}
\end{eqnarray}
\end{subequations}
then there exists a mapping
\begin{subequations}
\label{lin3}
\begin{eqnarray}
z_j&=&\phi_j(x,u),\;j=1,2,...,n\\\label{lin3a}
w^\gamma&=&\psi^\gamma(x,u),\;\gamma=1,2,...,m
\label{lin3b}
\end{eqnarray}
\end{subequations}
that transforms the given system to a linear system of PDEs. Here $F^\sigma$ is any arbitrary solution to the linear system of PDEs. In such a case, the components of the mapping $\phi_j$ and $\psi^\gamma$ satisfy the follwing set of PDEs:
\begin{subequations}
\label{lin4}
\begin{eqnarray}
\alpha_i^\sigma(x,u)\frac{\partial\Phi}{\partial x_i}
+\beta_\nu^\sigma(x,u)\frac{\partial\Phi}{\partial u^\nu}
&=&0,\;\;\;\;\;\sigma=1,2,...,m\\\label{lin4a}
\alpha_i^\sigma(x,u)\frac{\partial\Psi^\gamma}{\partial x_i}
+\beta_\nu^\sigma(x,u)\frac{\partial\Psi^\gamma}{\partial u^\nu}
&=&\delta^{\gamma\sigma},\;\;\sigma=1,2,...,m
\label{lin4b}
\end{eqnarray}
\end{subequations}
where $\delta^{\gamma\sigma}$ is a kronecker symbol, $\gamma,\sigma=1,2,...,m$.

In our case, the infinitesimal operator generating the infinite 
parameter group of 
transformation is 
\begin{eqnarray}
\label{lin5}
\hat{X} = X_c= c(t,x)(1-u)^2 \frac{\partial}{\partial u}.
\end{eqnarray}
From Eqs.~(\ref{lin4}) and (\ref{lin5}) we identify 
\begin{eqnarray}
\alpha_1^1=\alpha_2^1=0, \; \beta_1^1=(1-u)^2.
\label{in5a}
\end{eqnarray}
Then Eq.~(\ref{lin4}a) for $\Phi$ becomes
\begin{equation}
\Phi_u = 0.
\label{lin6}
\end{equation}
Clearly two independent solutions of Eq.~(\ref{lin6}) can be chosen as the new independent variables
\begin{equation}
z_1 = t, \;\;\; z_2 = x.
\label{lin7}
\end{equation}
Equation (\ref{lin4}b) for $\Psi$ becomes
\begin{equation}
(1-u)^2\Psi_u = 1.
\label{lin8}
\end{equation}
A particular independent solution of Eq.~(\ref{lin8}) can be easily found to be
\begin{equation}
w=\frac{1}{1-u}.
\label{lin9}
\end{equation}
As a result we obtain a mapping
\begin{eqnarray}
z_1=t, \quad z_2 = x, \quad w=\frac{1}{1-u}
\label{lin10}
\end{eqnarray}
which transforms the Fisher equation to a linear heat equation
\begin{eqnarray}
v_{z_1}-v_{z_2z_2}-v=0,\quad w=1+v.
\label{lin11}
\end{eqnarray}
Here we note that the above transformation is the same as the
B\"acklund transformation (\ref{pain10}). 
\section{The (2+1) Dimensional GF Equation }
\label{s5}
In general models that admit exact solutions are of considerable 
importance for understanding general behaviour of nonlinear 
dissipative systems. In one dimensions such models have received 
considerable importance. However, many realistic models are of 
two or three dimensional in nature and in this direction
Brazhnik and  Tyson [1999a,b] considered Eq.~(\ref{gfe2})
in two spatial dimensions and explored five kinds of travelling 
wave patterns, namely,  plane, V- and Y-waves, a separatrix and 
a space oscillating propagating structures. Thus,  it 
will be quite interesting to study the role of symmetries that 
allows the system to exhibit different spatiotemporal patterns 
and structures which usually will possess some kind of symmetry.
Moreover, the GF type equation does not lose its 
integrable nature (for $m=2$ case) on its dimension being 
increased. This is a 
very rare situation and so we consider 
the generalized Fisher equation in  2-spatial dimensions,
\begin{eqnarray}
\displaystyle{u_t-u_{xx}-u_{yy}-\frac{m}{1-u}(u_x^2+u_y^2)-u
+u^2=0},
\label{gf1}
\end{eqnarray}
and explore the underlying patterns. 

Extending the Lie symmetry analysis to Eq.~(\ref{gf1}) we find 
that the infinitesimal transformations
\begin{eqnarray} 
t &&\longrightarrow T =
t+\varepsilon\tau(t,x,y,u), \quad x \longrightarrow X =
x+\varepsilon\xi(t,x,y,u), \nonumber \\ 
y&& \longrightarrow Y =
y+\varepsilon\eta(t,x,y,u), \quad u \longrightarrow U =
u+\varepsilon\phi(t,x,y,u), \quad \varepsilon \ll 1,
\label{gf2}
\end{eqnarray}
again singles out the integrable case for  $m=2$ with the symmetries
\begin{eqnarray}
\tau=a,
\quad \xi=by+l, \quad \eta = -bx+n,\quad \phi = c(t,x,y)(1-u)^2,
\label{gf3}
\end{eqnarray} 
where $ c(t,x,y) $ is any solution to the two dimensional
linear heat  equation
$c_t-c_{xx}-c_{yy}-c = 0$ and $b, l$ and $n$ are arbitrary constants. 
For other values of $m( \ne 2)$, we get the following symmetries
\begin{eqnarray} 
\tau=a, \quad
\xi=by+l, \quad \eta = -bx+n, \quad \phi = 0.
\label{gf4}
\end{eqnarray} 

\subsection{\emph{ \bfseries Lie algebras}}
The symmetry algebra of the (2+1) dimensional Fisher type 
equation (\ref{gf1}) with $m=2$
is spanned by four vector fields  
\begin{equation}
X_1 = \partial_t,\;\;
X_2 = y\partial_x-x\partial_y, \;\;
X_3 = \partial_x, \;\;
X_4 = \partial_y, \label{gf5}
\end{equation}
and a vector field associated with the infinite-dimensional subalgebra
\begin{equation}
 X_c = c(t,x,y) (1-u)^2\partial_u.
\label{gf6}
\end{equation}
The physical interpretation of the vector fields is the following.  Vector 
field $X_1$ indicates that (\ref{gf1}) is invariant under time translation 
whereas vector fields $X_3$ and $X_4$ reflect the spatial invariance of 
(\ref{gf1}) in $x$ and $y$ directions respectively.  Vector field $X_2$ 
demonstrates that Eq.~(\ref{gf1}) is invariant under rotation.  

The commutation relations between these vector fields become
\begin{eqnarray}
&&[X_1,X_2]=0,\quad \;\;[X_1,X_3]=0,\quad \;\;\;\;[X_1,X_4]=0,
\qquad \quad [X_1,X_c]=X_{c_t}, 
\nonumber\\ &&[X_2,X_3]= X_4,
\;\; \;[X_2,X_4]=-X_3,\;\;[X_2,X_c]=yX_{c_x}-xX_{c_y},\nonumber
\\
&&[X_3,X_4]=0,\quad \;\;[X_3,X_c]=X_{c_x},\;\;\;\;[X_4,X_c]=X_{c_y},
\label{gf7}
\end{eqnarray}
where $X_{c_t} = c_t(1-u)^2\partial_u, \quad X_{c_x} = c_x(1-u)^2
\partial_u, \quad \mbox{and} \quad 
X_{c_y} = c_y(1-u)^2\partial_u.$ 

Proceeding in  a similar fashion for the
nonintegrable case $(m\ne2)$ we obtain the associated Lie algebra as
\begin{eqnarray}
&&[X_1,X_2]=0,\quad \quad [X_1,X_3]=0,\qquad \;\; [X_1,X_4]=0, \nonumber
\\
&&[X_2,X_3]= X_4,\quad \;[X_2,X_4]=-X_3,\quad [X_3,X_4]=0,
\label{gf8}
\end{eqnarray}
which is obviously a subcase of (\ref{gf7}).
\section{Similarity Solutions of the (2+1) Dimensional Integrable 
GF Equation}
\label{s6}
Here we solve the characteristic equation
\begin{eqnarray}
\displaystyle
{\frac {dt} {\tau} \quad = \quad \frac {dx}{\xi} = \quad \frac {dy}{\eta} 
\quad = \quad
\frac{du} {\phi}}
\label{gf9}
\end{eqnarray} 
and obtain the similarity reduced PDE in two independent 
variables. The resultant PDE is then analysed for its symmetry 
properties to reduce it to an ODE so that explicit solutions 
can be found by integrating it. In the general reduction, we 
obtain cylindrical function solutions. We derive certain 
physically important structures by restricting some of the 
symmetry parameters to be zero. For example, we obtain Bessel 
function solutions by assuming one of the symmetry parameters to be zero. Similarly proceeding we report certain 
propagating type  structures and so on by restricting appropriately the choice of the symmetry parameters
involved. In such a process, we report all the 
propagating structures discussed by Brazhnik and  
Tyson [1999a,b] as a subcase of our results. Finally, we 
put forth the relation between symmetries of the GF 
equation and the 2-dimensional linear heat equation. 
\subsection{\bf \emph{ General reductions}}
Now integrating Eq.~(\ref{gf9}) with the infinitesimal symmetries 
(\ref{gf3}) along with the condition $b\ne0$ (as well as $l,n\ne0$), 
we obtain  the following 
similarity variables:
$$ 
\displaystyle{z_1=\frac{b}{2}(x^2+y^2)+ly-nx,\quad z_2=-t
-\frac{a}{b}\sin^{-1}\left(\frac{n-bx}{\sqrt{l^2+n^2+2bz_1}}
\right),} 
\nonumber 
$$
\begin{equation}
u=1- \displaystyle { \frac {a}{w(z_1,z_2)+\int c(t,x,y)dt}}.
\label{gr1}
\end{equation}
In this case also, one can obtain the 
linear heat equation
\begin{equation}
\chi_t-\chi_{xx}-\chi_{yy}-\chi=0
\label{gf11}
\end{equation}
by assuming $\chi=\displaystyle{\frac{1}{a}}\Big(w(z_1,z_2)+\int c(t,x,y)dt
\Big)$  in (\ref{gr1}) and substituing the resultant transformation into Eq.~(\ref{gf1}). Then  
the \emph{similarity reduced variables} (\ref{gr1}) can be
interpreted
as giving rise to the \emph{linearizing transformation} from the 
group theoretical point of view. 
Under the above similarity transformations, Eq.~(\ref{gf1}) gets reduced to
a PDE in two independent variables $z_1$ and $z_2$,
\begin{eqnarray}
\displaystyle{w_{z_2}+2bw_{z_1}+(2bz_1+l^2+n^2)w_{z_1z_1}+\frac {a^2w_
{z_2z_2}}{2bz_1+l^2+n^2}+w-a=0}.
\label{gr2}
\end{eqnarray}

The similarity reduced PDE (\ref{gr2}) can itself be further 
analysed for the symmetry properties by once again subjecting it 
to the invariance analysis under the classical Lie algorithm.
For this purpose we first redefine  the variables as 
\begin{equation}
\bar {z_1} = 2bz_1+l^2+n^2,\quad \bar {z}_2 = z_2, \quad
\bar{w}=w-a,\quad b,n \ne 0
\label{gr3}
\end{equation}
so that Eq.~(\ref{gr2}) becomes
\begin{equation}
\bar{w}_{\bar{z}_2}+4b^2\bar{w}_{\bar{z}_1}
+4b^2\bar{z}_1\bar{w}_{\bar{z}_1\bar{z}_1}
+\displaystyle{\frac{a^2}{\bar{z}_1}}\bar{w}_{\bar{z}_2 \bar{z}_2}+\bar{w}=0.
\label{gr4}
\end{equation}
Eq.~(\ref{gr4}) can be shown to possess the following symmetries
\begin{equation} 
\tau = c_3,
\quad \xi = 0, \quad \phi=c_1\bar {w}+g(\bar {z}_1, \bar
{z}_2),
\label{gr5}
\end{equation}
where $ c_1 $ and
$ c_3 $ are arbitrary constants and $ g(\bar{z_1},\bar{z_2})$,   
which arises due to the linear nature of Eq.~(\ref{gr4}), is any 
solution to the equation
$g_{\bar{z}_2} + 4b^2g_{\bar{z}_1} + 4b^2\bar {z}_1
g_{\bar{z}_1\bar{z}_1}+\displaystyle{\frac{a^2}{\bar {z}_1}}
g_{\bar{z}_2\bar{z}_2}+g=0.$
The corresponding Lie vector fields 
\begin{equation}
X_1=\partial_{\bar{z}_2}, \quad X_2=\bar{w}\partial_{\bar{w}}, \quad
X_g=g(\bar{z}_1,\bar{z}_2) \partial_{\bar{w}}
\label{gr6}
\end{equation}
lead to the infinite-dimensional Lie algebra
\begin{equation}
[X_1,X_2]=0,\quad [X_1,X_{g}]=X_{g_{z_2}},\quad
[X_2,X_{g}]=-X_{g}, 
\label{gr7}
\end{equation}
where $X_{g_{z_2}}=g_{z_2}(\bar{z_1},\bar{z_2})\partial_{\bar{w}}$.
By solving the characteristic equation associated with the symmetries 
(\ref{gr5}) we obtain new similarity variables 
\begin{equation} 
\zeta=\bar {z}_1,
\quad {w}= a+e^{ \left(\displaystyle{ \frac {c_1\bar {z}_2}{c_3}}\right
)}\left[f(\zeta)+\displaystyle{\frac{1}{c_3} \int g(\bar {z}_1, 
\bar {z}_2)} e^{\displaystyle{ \left
(-{\frac {c_1}{c_3}} \bar {z}_2 \right)}} d\bar {z}_2  
\right], 
\label{gr8}
\end{equation}
%
where $f(\zeta)$ satisfies the linear second order ODE of the form 
\begin{eqnarray} 
\zeta^2 f''+\zeta f'+(A+B \zeta)f=0,\quad A=
(ac_1/2bc_3)^2, \quad B = (1+c_1/c_3)/4b^2,
\label{gr9}
\end{eqnarray} 
and prime  denotes differentiation with respect to $\zeta $. The
exact solution to Eq.~(\ref{gr9}) can be expressed in terms of cylindrical 
functions of the form [Murphy, 1969]
\begin{eqnarray} 
f=I_1Z_1(2\sqrt{B\zeta})+I_2Z_2(2\sqrt{B\zeta}), 
\label{gr10}
\end{eqnarray} 
where $ Z_i(\zeta) $, $ i=1,2 $, are two
linearly independent cylindrical functions and $I_1$, $I_2$ are
arbitrary constants. Then the invariant solution to the (2+1)-dimensional PDE 
(\ref{gf1}) can be written as 
\begin{eqnarray} 
&u = 1- a \Bigg[ a + e^{\displaystyle {\left(
{\frac{c_1}{c_3}} \bar{z}_2\right)}} \Bigg (I_1Z_1(2\sqrt{B\bar{z}_1})+ 
I_2Z_2(2\sqrt{B\bar{z}_1})
-\frac{1}{c_3}\int g(\bar{z}_1,\bar{z}_2)
e^{\displaystyle{\left({\frac{c_1}{c_3}\bar{z}_2}\right)}} 
d\bar{z}_2\Bigg) \Bigg.&  
\nonumber \\   &\mskip 480 mu \Bigg.
 + \displaystyle{\int c(t,x,y)dt} \Bigg]^{-1},&
\label{gr11}
\end{eqnarray}
where $\bar{z}_1$ and $\bar{z}_2$ are of the form (\ref{gr1}) via
(\ref{gr3}).
Here $g(\bar{z}_1,\bar{z}_2)$ can also be taken as zero. However, 
the solution (\ref{gr11}) is richer in structure
when $g(\bar{z}_1,\bar{z}_2)$ is taken as non-zero.

\subsection{\bfseries\emph{{Special Reductions}}}
Besides the above general solution,  some physically interesting 
structures 
can be obtained  by choosing some 
of the symmetry parameters including the arbitrary functions 
to be zero. In the following we present certain nontrivial cases. Particularly, we consider separately the two special cases: (a) $c_1=0$ and (b)$c_1=-c_3$.

\subsubsection*{{6.2(A) \bf \emph{The choice $b\ne0$}}}
{\bfseries Case (a): $c_1=0$}.

Solving the characteristic equation associated with (\ref{gr5})
with $c_1=0$ we obtain the similarity variables
\begin{eqnarray} 
\zeta=\bar{z}_1,
\quad {w}= a+f(\zeta)+\displaystyle{\frac{1}{c_3} \int g({z}_1, 
 {z}_2)}d{z}_2.
\label{gr12}
\end{eqnarray} 
Here the function $f(\zeta)$ satisfies the ODE
\begin{eqnarray}
\zeta f''+f'+Bf=0, \quad B=\displaystyle{\frac{1}{4b^2}},\;\; 
'=\frac{d}{d\zeta}.
\label{gr13}
\end{eqnarray}
The solution (static) to (\ref{gr13}) is found to be
\begin{eqnarray}
f(\zeta)=I_1J_0(2\sqrt{B\zeta})+I_2Y_0(2\sqrt{B\zeta}),
\label{gr14}
\end{eqnarray}
where $J_0$ and $Y_0$ are zeroth order Bessel functions of first and 
second kinds respectively.
Then the time-independent solution to the original PDE (\ref{gf1}) becomes 
\begin{eqnarray}
u=&&1-a\Bigg[a+I_1J_0\Big(2\sqrt{B\left(b^2(x^2+y^2)+2b(ly-nx)+l^2
+n^2\right)}\Big)\Bigg. \Big. \nonumber\\ \Big. \Bigg. 
&& \;\;\;\;\;\;\;\; +I_2Y_0\Big(2\sqrt{B\left(b^2(x^2+y^2)+2b(ly-nx) 
+l^2+n^2\right)}\Big) \Bigg.\label{gr15} \\\Bigg.
&& \;\;\;\;\;\;\;\; +\displaystyle{\frac{1}{c_3}}
\int g(\bar{z}_1,\bar{z}_2)d\bar{z}_2
+\int c(t,x,y)dt\Bigg]^{-1}.
\nonumber
\end{eqnarray}

The solution plot is given in Figs.~(5) for $g(\bar{z}_1, \bar{z}_2)=0$ and
$c(t,x,y)=0$. For the choice  $I_1\ne0,I_2=0$, Fig.~(5a) shows a  static
circular symmetric pattern. If we choose  $I_2\ne0$ while keeping   $I_1=0$  we
get a static structure (Fig.~(5b)) exhibiting circular symmetry with a 
singularity at the origin and it is due to the nature of Bessel function of
second  kind. Finally  for $I_1, I_2\ne0$ the plot is shown in  Fig.~(5c). 
\begin{figure}[!ht]
\begin{center}
\epsfig{file=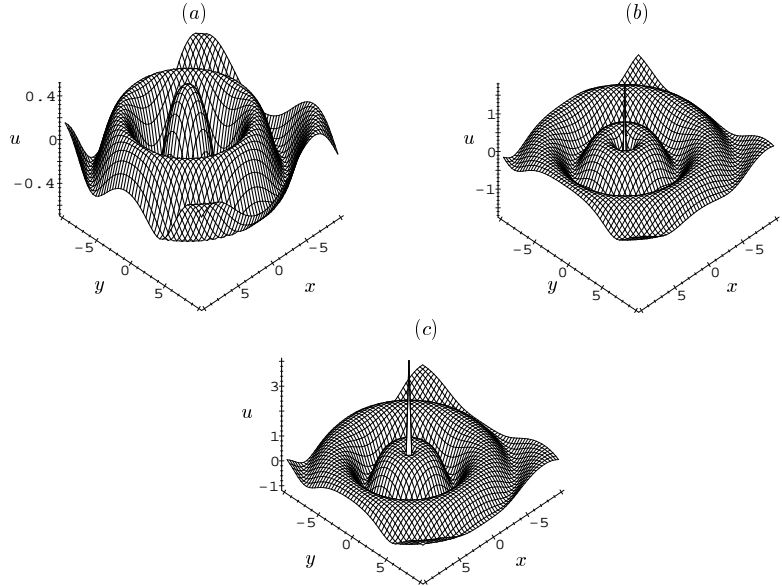,width=.75\linewidth}
\end{center}
\caption{ Static structures exhibiting (\ref{gr15}) circular symmetry:  
(a) $I_1\ne0, I_2=0$ (b) $I_1=0,  I_2\ne0$ (c) $I_1, I_2\ne0$.}
\label{static}
\end{figure}

{\bfseries Case (b): $c_1=-c_3$}.

Making use of the choice $c_1=-c_3$ in (\ref{gr8}) we obtain
the  similarity variables
\begin{eqnarray} 
\zeta=\bar{z}_1,
\quad {w}= a+e^{-\bar{z}_2}\Bigg[f(\zeta)+\displaystyle{\frac{1}{c_3} 
\int g({z}_1,{z}_2)} e^{\bar{z}_2}d{z}_2\Bigg].
\label{gr16}
\end{eqnarray} 
These similarity transformations reduce the PDE (\ref{gr4})
to an ODE 
\begin{eqnarray}
\zeta^2 f''+\zeta f'+Af=0, \quad A=\left(\displaystyle{\frac{a}{2b}}\right)^2,
\label{gr17}
\end{eqnarray}
where prime denotes differentiation with respect to $\zeta$.  
Solving Eq.~(\ref{gr17}) we find 
\begin{eqnarray}
f=I_1\zeta^{m_1}+I_2\zeta^{m_2}, \quad m_{1,2}=\pm \sqrt{-A}.
\label{gr18}
\end{eqnarray}
Since $A>0$, bounded  solution can be written as 
\begin{eqnarray} 
u=&&1- a \Bigg [ a + e^{\displaystyle {-z_2}}
\Bigg (I_1\Bigg|\cos\Big(\sqrt{A}\log\left|\left(b^2(x^2+y^2)+2b(ly-nx)+l^2
+n^2\right)\right|\Big)\Bigg|\Bigg.
\Bigg. \nonumber\\ \Bigg. \Bigg.&&+ 
I_2\Bigg|\sin\Big(\sqrt{A}\log\left|\left(b^2(x^2+y^2)+2b(ly-nx)
+l^2+n^2\right)\right|\Big)\Bigg|\Bigg)\Bigg]^{-1}.
\label{gr19}
\end{eqnarray}
Here the variable $z_2$ is given by Eq.~(\ref{gr1}).  The solution plot 
is given in Figs.~(6) for $g(z_1,z_2)$, $c(t,x,y) = 0$
and it exhibits a complicated propagating pattern. 
\begin{figure}[!ht]
\begin{center}
\epsfig{file=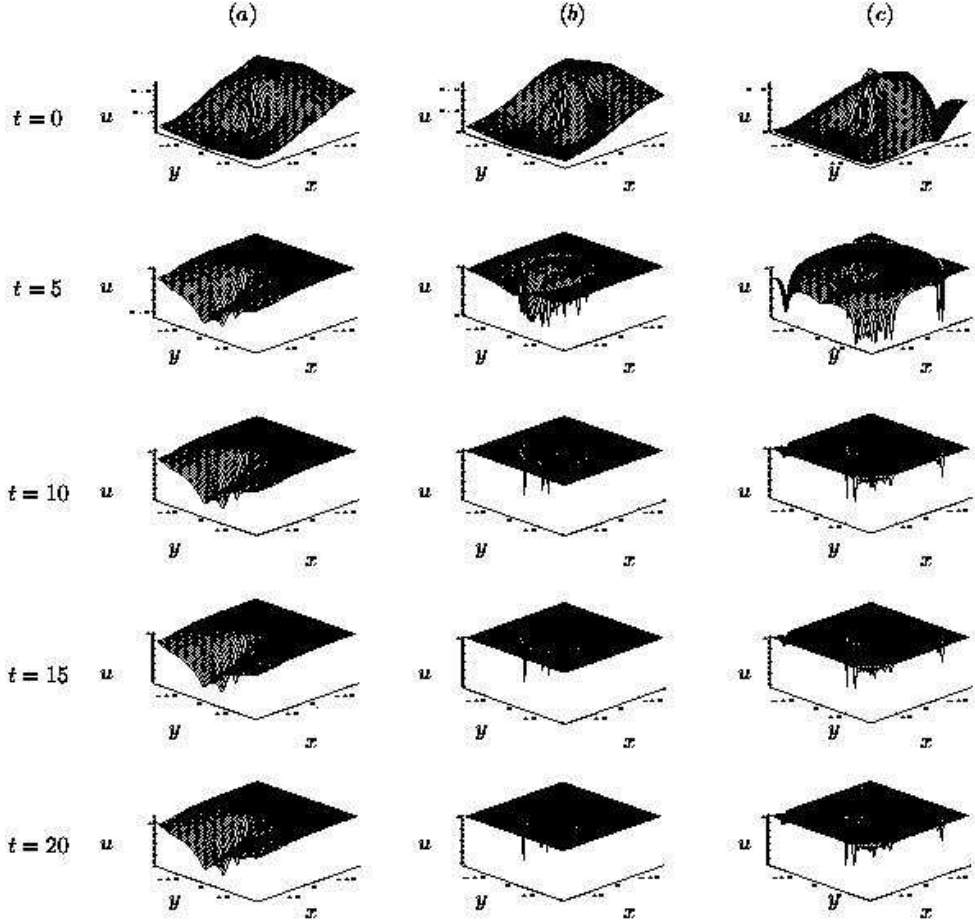,width=.9\linewidth}
\end{center}
\caption{ Propagating  type structures corresponding to Eq.~(\ref{gr19})
at various instants:  (a) $I_1,I_2\ne0$ (b) $I_1\ne0,  I_2=0$ (c) $I_1=0, I_2\ne0$.}
\label{cosif}
\end{figure}

\subsubsection*{6.2(B) \bf \emph{The choice $b=0$}}
In the general reduction,  we have assumed $b\ne0$.  
In this subsection, we derive certain interesting patterns by taking  
the symmetry parameter $b=0$.  For this choice 
the symmetries are (vide Eq.~(\ref{gf3}))
\begin{eqnarray} 
\tau=a, \quad \xi=l, \quad \eta = n,\quad \phi =c(t,x,y)(1-u)^2. 
\label{sr1}
\end{eqnarray} 
The similarity transformations 
\begin{equation} 
z_1=nx-ly,\;\;z_2=ax-lt,\;\;
u=1-\displaystyle{\frac {a}{w(z_1,z_2)+\int c(t,x,y)dt}},    
\label{sr2}                        
\end{equation}
are found by solving the characteristic equation associated with 
Eq.~(\ref{sr1}).
Under the above set of transformations (\ref{sr2}), Eq.~(\ref{gf1}) 
can be transformed to 
\begin{eqnarray}
lw_{z_2}+(l^2+n^2) w_{z_1z_1}+a^2 w_{z_2z_2}
+2anw_{z_1z_2}+w-a=0.
\label{sr3}
\end{eqnarray}
The invariance of Eq.~(\ref{sr3}) under the one-parameter Lie
group transformations leads to the infinitesimals
\begin{eqnarray}
& \tau= 2a\displaystyle{\frac{n}{l}}c_1z_2-2
\displaystyle{\frac{a^2}{l}}c_1z_1+c_2, \quad
\xi = -2a\displaystyle{\frac{n}{l}}c_1z_1
+\displaystyle{\frac{2}{l}}(l^2+n^2)c_1z_2-c_3, &\nonumber\\
& \phi= (c_1z_1+c_4)\bar{w}+h(z_1,z_2),&
\label{sr4}
\end{eqnarray}
where $ \bar{w}=w-a $ and the arbitrary function 
$ h(z_1,z_2)$ satisfies the equation
$lh_{z_2}+(l^2+n^2)h_{z_1z_1}+a^2h_{z_2z_2}
+2anh_{z_1z_2}$ $+h=0.$

Using the symmetries given in Eq.~(\ref{sr4}) one can construct a general 
similarity reduced ODE.  However, as our aim is to show some interesting 
solutions we consider a lesser symmetry parameter group in the 
following. 

{\bfseries Case (a)}: 
To begin with let us consider $c_1\ne0$ and 
 all other parameters $(c_2,c_3,c_4)$ are zero in (\ref{sr4}).  Now solving 
the characteristic equation associated with the symmetries 
we obtain the following similarity variables 
\begin{eqnarray}
\zeta & = & \frac{a^2}{2}z_1^2+\frac{(l^2+n^2)}{2}z_2^2
-anz_1z_2,\nonumber\\
\bar{w} & = & e^{-\displaystyle{F(\zeta,z_1)}}
\Bigg[f(\zeta)-\displaystyle{\frac{(al)^2}{c_1}}
\displaystyle{\int F(\zeta,z_1)h(z_1,z_2)
e^{\displaystyle{F(\zeta,z_1)}}dz_1}\Bigg],
\label{sr5}
\end{eqnarray}
where $
F(\zeta,z_1) = -\displaystyle{\frac{1}{2a^2l}
\Big[2(l^2+n^2)\zeta-a^2l^2z_1^2
\Big]^{\frac{1}{2}}}$.
Using the above similarity transformations we deduce an ODE
\begin{eqnarray}
\zeta f''+f'+Bf=0, \quad B=\displaystyle{\frac{1}{2a^2l}}\Bigg(1-
\displaystyle{\frac{l^2+n^2}{4a^2}}\Bigg),
\label{sr6}
\end{eqnarray}
where the prime stands for differentiation with respect 
to $\zeta$. 
Eq.~(\ref{sr6}) admits a  time-dependent cylindrical
wave solution of the form 
\begin{eqnarray}
f(\zeta)=I_1J_0(2\sqrt{B\zeta})+I_2Y_0(2\sqrt{B\zeta}).
\label{sr7}
\end{eqnarray}
From (\ref{sr7})  we can provide the time-dependent solution to the 
original PDE through the transformations (\ref{sr5}) and (\ref{sr2}).
The solution is plotted in Figs.~(7) for various instants,
exhibiting propagating spike like patterns.
\begin{figure}[!ht]
\begin{center}
\epsfig{file=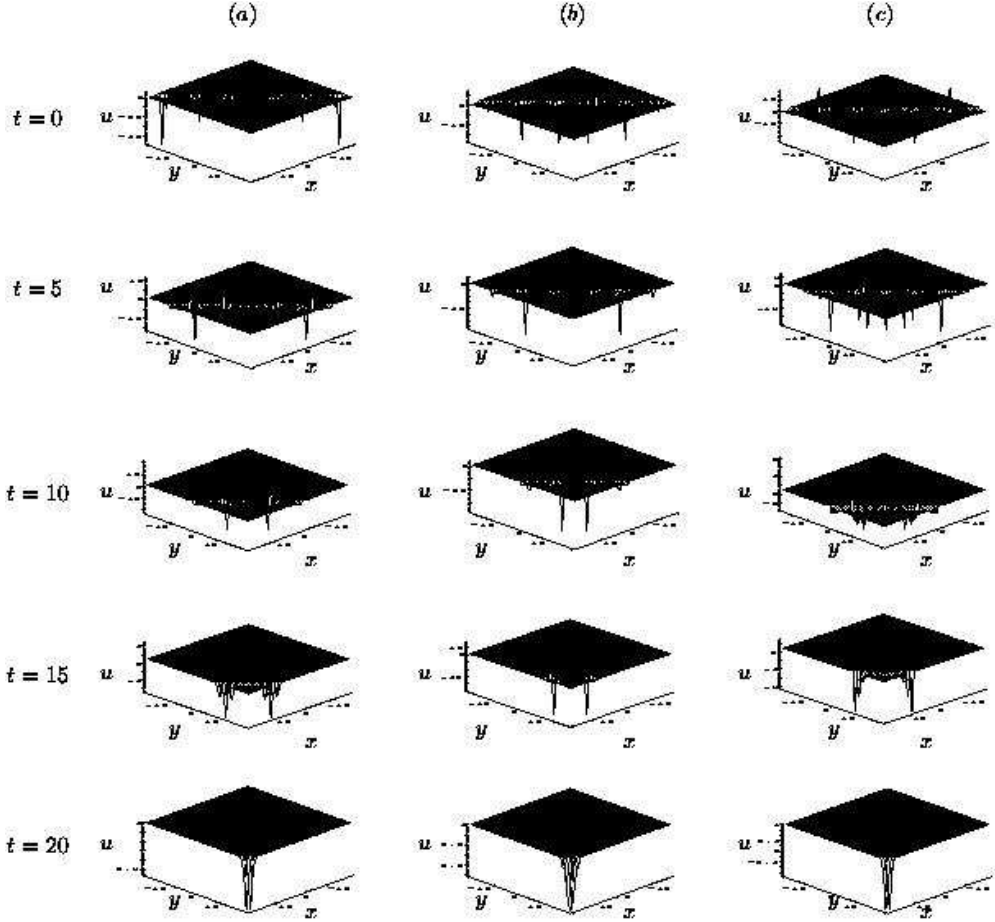,width=.9\linewidth}
\end{center}
\caption{ Bessel type propagating solutions for various instants as given by 
Eq.~(\ref{sr7}):  (a) $I_1,I_2\ne0$, (b) $I_1\ne0,  I_2=0$ (c) $I_1=0, I_2\ne0$.}
\label{bessel}
\end{figure}

{\bfseries Case (b): $c_1=0$}.

The choice  $c_1=0$ and $c_2,c_3,c_4 \ne 0$ in Eq.~(\ref{sr4}) leads to the similarity
transformations 
\begin{eqnarray}
\zeta=c_3z_2+c_2z_1,\quad
\bar{w}=e^{\displaystyle{kz_2}}\left[f(\zeta)+\frac{1}{c_2}\int h(z_1,z_2)e^{\displaystyle{k}z_2}dz_2\right],\;k=\frac{c_4}{c_2}
\label{ssr7}
\end{eqnarray}
which transforms Eq.~(\ref{sr3}) into an ODE of the form
\begin{eqnarray}
&Af''+Bf'+Cf=0,\quad  A=(ac_3)^2+(l^2+n^2)c_2^2+2anc_2c_3, &\nonumber \\
&B=lc_3+2a^2kc_3+2anc_4,\quad  
C=1+lk+(ak)^2,\;\; '=\frac{\displaystyle{d}}
{\displaystyle{d\zeta}},&
\label{sr8}
\end{eqnarray}
whose general solution is given by
\begin{eqnarray}
f(\zeta)=I_1e^{\displaystyle{m_1\zeta}}
+I_2e^{\displaystyle{m_2\zeta}},\quad 
m_{1,2}=-\frac{1}{2A}\left(B\pm\sqrt{B^2-4AC}\right),
\label{sr9}
\end{eqnarray}
and $\zeta$ is quadratic in time $t$.
Substituting the similarity transformations  (\ref{ssr7}) and  
(\ref{sr2}) in (\ref{sr9}) the solution to the original
PDE (\ref{gf1}) can be obtained.

{\bfseries Case (c): {Propagating structures}}.

By assuming $n=c_3=0$ in Eq.~(\ref{sr8}), the ODE becomes
\begin{eqnarray}
f''-k_1f=0, \quad k_1=-\displaystyle{\frac{1+lk+(ak)^2}{(lc_2)^2}}.
\label{sr10}
\end{eqnarray}
Then, the system (\ref{gf1}) is found to exhibit  
propagating structures and the  corresponding solutions are given by
\begin{eqnarray}
u=\left\{
	\begin{array}{lll}
1-a\Bigg \{a+\exp\Big [\displaystyle{{k\left(ax-lt\right)}}\Big ]
\Bigg [ I_1\cos(\sqrt{k_1}c_2ly)+I_2\sin(\sqrt{k_1}c_2ly) \Bigg . \Bigg .\\ 
\Bigg. \Bigg.
\mskip 150 mu+\displaystyle{\int \frac{h(z_1,z_2)}{c_5}}
e^{\displaystyle {kz_2}}dz_2 \Bigg ] 
+\displaystyle {\int c(t,x,y)dt}\Bigg \}^{-1}, \quad \quad k_1<0; \\ [2mm]
1-a\Bigg \{a+\exp\Big [\displaystyle {k\left(ax-lt \right)}
\Big ] 
\Bigg [ I_1e^{\displaystyle{\sqrt{k_1}c_2ly}}
+I_2e^{\displaystyle{-\sqrt{k_1}c_2ly}}\Bigg . \Bigg. \\
\Bigg. \Bigg. 
\mskip 150 mu +\displaystyle{\int \frac{h(z_1,z_2)}{c_5}}
e^{\displaystyle{kz_2}}dz_2 \Bigg ]
+\displaystyle{\int c(t,x,y)dt}\Bigg \}^{-1}, \quad \quad k_1>0; \\[2mm]
1-a\Bigg \{a+\exp\Big[\displaystyle{k\left(ax-lt \right)}\Big ] 
\Bigg [ I_1c_2ly+I_2 \Bigg. \Bigg. \\ \Bigg. \Bigg. \mskip 150 mu
+\displaystyle{\int \frac{h(z_1,z_2)}{c_5}}e^{\displaystyle{kz_2}}dz_2\Bigg ]
+\displaystyle{\int c(t,x,y)dt} \Bigg \}^{-1}, \quad \quad k_1=0. \\[2mm]
\end{array}
\right. 
\label{sr11}
\end{eqnarray}
The presence of an additional function $h(z_1,z_2)$ in the solution 
(\ref{sr11}) leads to a more general form. In particular, Eq.~(\ref{sr11})
exhibits the five classes of 
bounded travelling wave solutions reported by Brazhnik and Tyson 
[1999a] with specific choice
of the parameters involved along with $h(z_1,z_2) = 0$ and $c(t,x,y) = 0$. 
The corresponding  solutions are given below.

The simplest travelling wave solution (Fig.~8a)
\begin{eqnarray}
u=1-\displaystyle{\frac{1}{1 + A \exp \left[k\left(\displaystyle
{ax-lt}\right)\pm\sqrt{k_1}c_2ly \right]}}, 
\quad \quad k_1>0, \quad A=\displaystyle{\frac{I_2}{a}}\; \mbox{or}\;
\displaystyle{\frac{I_1}{a}}
\label{gf33}
\end{eqnarray}
is constructed by assuming either $I_1 = 0$ or $ I_2 = 0$.
When $I_1=I_2(\ne0)$, one obtains a V-wave pattern (Fig.~8b)
\begin{eqnarray}
u=1-\displaystyle{\frac{1}{1 + A \exp \left[\displaystyle{k\left(
ax
-lt\right)}\right]\cosh(\sqrt{k_1}c_2ly)}}, 
\; \; k_1>0, \quad A=\displaystyle{\frac{2}{a}}.
\label{gf34}
\end{eqnarray}
A wavefront oscillating in space (Fig.~8c) 
\begin{eqnarray}
u=1-\displaystyle{\frac{1}{1 + A \exp \left[\displaystyle{k\left(ax-lt\right)}\right]|\cos(\sqrt{k_1}c_2ly)|}},\quad A=\displaystyle{\frac{I_1}{a}}
\label{gf35}
\end{eqnarray}
exists for the choice $I_2=0$ and $k_1 < 0$.
Further for $I_1\ne 0$ and $I_2=0$ we get a separatrix 
solution (Fig.~8d)
\begin{eqnarray}
u=1-\displaystyle{\frac{1}{1 + A |y|\exp \left[\displaystyle{k\left(ax-lt\right)}\right]}},\quad k_1=0,\quad A=\displaystyle{\frac{I_1lc_2}{a}}.
\label{gf36}
\end{eqnarray}
 Finally,  for positive $k_1$ with $ I_1=-I_2 $ the Y-wave solution 
(Fig.~8e)
\begin{eqnarray}
u=1-\displaystyle{\frac{1}{1 + A \exp\left[\displaystyle{k\left(ax-lt\right)}\right]
|\sinh(\sqrt{k_1}c_2ly)|}}, \quad A=\displaystyle{\frac{I_1}{a}}
\label{gf37}
\end{eqnarray}
exists. In each of  the above solutions $A$ is a positive constant.
As Fisher equation forms a basis for many nonlinear models, the above 
solutions are nothing but reminiscent of patterns from different fields.
To mention a few, V-waves are characterized in the framework of 
geometrical crystal growth related models [Schwendeman, 1996] and in
excitable media [Brazhnik \& Davydov, 1995], while space oscillating fronts are 
relevant to cellular flame  structures and patterns in chemical 
reaction diffusion 
systems [Scott  \& Showalter, 1992; Showalter, 1995]. Besides, excitable media supporting space-oscillating
fronts are discussed by Brazhnik \emph{et al.} [1996] with a 
geometrical model.
\begin{figure}[!ht]
\begin{center}
\epsfig{file=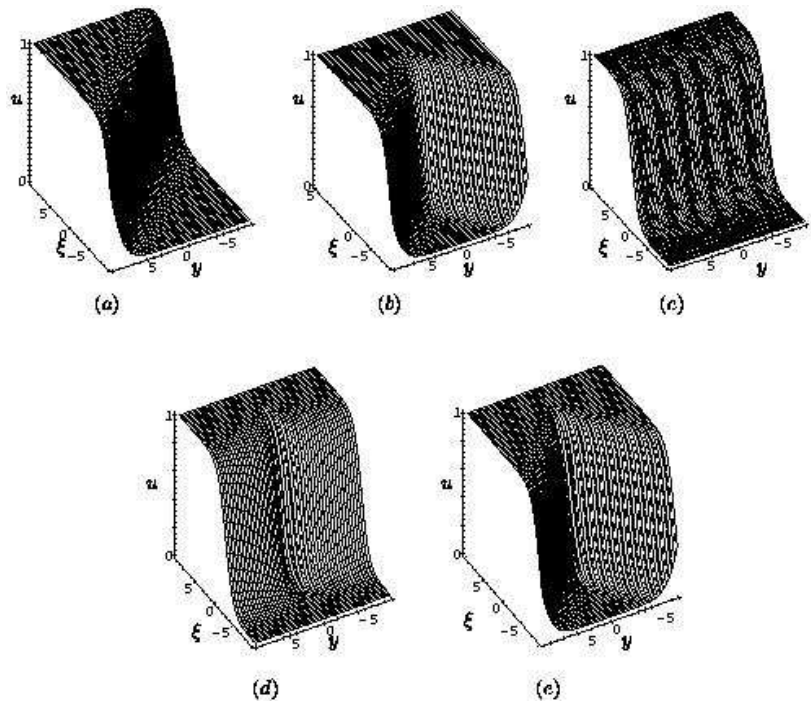,width=.75\linewidth}
\end{center}
\caption{  Five interesting classes of propagating wave patterns as obtained
in ref.[Brazhnik \& Tyson, 1999a], which follow from Eqs.~(\ref{sr11}): (a) Travelling 
waves, (b) V-waves, (c) Oscillating front, (d) Separatrix solution, 
(e) Y-waves with $\xi=k(ax-lt)$.}
\label{waves}
\end{figure}
 
\subsection{\bfseries\emph{Relation between symmetries of $m=2$ case and of linear 
heat equation in 2-spatial dimensions}}
Proceeding in a similar fashion as that for the (1+1) dimensional 
Fisher equation as given in subsection \ref{sec4c} one 
obtains a mapping
\begin{eqnarray}
z_1=t,\;z_2=x,\;z_3=y,\; w=\frac{1}{1-u}
\label{gf75}
\end{eqnarray}
by solving Eqs.~(\ref{lin4}) for the two spatial dimensional 
case. 
Eq.~(\ref{gf75}) then transforms the Fisher equation in 2-spatial 
dimensions to a 2-dimensional linear heat equation
\begin{eqnarray}
v_{z_1}-v_{z_2z_2}-v_{z_3z_3}-v=0,\; w=1+v.
\label{gf76}
\end{eqnarray}
However, as in the case of the (1+1) dimensional GF equation,
 we have 
given an interpretation for the transformation 
(\ref{gf75}) in terms of the B\"acklund transformation via Eq.~(\ref{pain10}). 
\section{ The Nonintegrable  (2+1) Dimensional GF Equation}
\label{s7}
One can make use of the same techniques that used for the integrable 
case
to obtain some special solutions for the nonintegrable case ($m\ne2$). 
To start with, we shall consider the general 
reductions. Due to the nonintegrable nature of the equation under 
consideration, the general reductions lead  to  ODEs of
non-Painlev\'e type. However, we obtain certain special 
solutions, like plane wave solutions, static structures 
and localized
structures under appropriate parametric restrictions.
\subsection{\bfseries\emph{General reductions}}
To begin with we derive a general similarity reduced ODE by assuming 
none of the parameters in (\ref{gf4}) to be zero.
The  similarity variables thus obtained are
\begin{eqnarray} 
& \displaystyle{z_1=\frac{b}{2}(x^2+y^2)+ly-nx,\quad z_2=-t
-\frac{a}{b}\sin^{-1}\left(\frac{n-bx}{\sqrt{l^2+n^2+2bz_1}}
\right),}  &\nonumber\\
 & u=w(z_1,z_2).&
\label{gf45}
\end{eqnarray}
Under the transformations (\ref{gf45}) the reduced PDE in two independent 
variables takes the form
\begin{eqnarray}
\displaystyle{w_{z_2}+2bw_{z_1}+(2bz_1+l^2+n^2)w_{z_1z_1}+\frac{a^2
w_{z_2z_2}}
{2bz_1+l^2+n^2}} +\displaystyle{\frac{m}{1-w}}\nonumber\\ \times
\displaystyle{\left[(2bz_1+l^2
+n^2)w_{z_1}^2+\frac{a^2w_{z_2}^2}{2bz_1+l^2+n^2}\right]
-w+w^2}& =  0. 
\label{gf46}
\end{eqnarray} 
Again looking  the invariance properties of Eq.~(\ref{gf46}) we get the
following infinitesimals:
\begin{eqnarray}
\tau=c_1,\quad \xi=0,\quad \phi=0. 
\label{gf47}
\end{eqnarray} 
These infinitesimals reduce Eq.~(\ref{gf46}) to an ODE 
\begin{eqnarray}
4b^2\left(\zeta f''+f'+\frac
{m\zeta}{1-f}f^{'2}\right)+f-f^2=0
\label{gf48}
\end{eqnarray}
under a set of similarity transformations
\begin{eqnarray}
\zeta=\bar{z}_1, \quad w=f(\zeta), \quad \bar{z}_1=2bz_1+l^2+n^2.
\label{gf49}
\end{eqnarray}
As Eq.~(\ref{gf48}) is not integrable, we look for special reductions 
with lesser number of parameters involved. 

\subsection{\bfseries\emph{Special reductions}}
For the special case $b=0$ the infinitesimals obtained are
$\tau=a,\xi=l,\;\eta = n,\;\phi = 0$.  
The associated similarity transformations $z_1=nx-ly,\;
z_2=ax-lt,\;u=w(z_1,z_2)$ reduces the Fisher type equation (\ref{gf1}) 
to a PDE
\begin{eqnarray}
lw_{z_2}+(l^2+n^2)w_{z_1z_1}+a^2w_{z_2z_2}+2anw_{z_1z_2}
+\frac{m}{1-w}\nonumber\\  \times
\displaystyle{\left[(l^2+n^2)w_{z_1}^2+a^2w_{z_2}^2+2anw_{z_1}w_{z_2
}\right] +w-w^2} & =0
\label{gf59}
\end{eqnarray} 
in two independent variables $z_1$ and $z_2$. Applying classical Lie 
algorithm on Eq.~(\ref{gf59}) one obtains the following infinitesimals
\begin{eqnarray} 
\tau=c_2,\quad \xi =
\frac{n(c_1+c_2)}{a}, \quad \phi = 0. 
\label{gf60}
\end{eqnarray} 
The corresponding
similarity variables, $\zeta=-ac_2z_1+n(c_1+c_2)z_2$ and $w=f(\zeta)$, 
reduce Eq.~(\ref{gf59}) to
\begin{eqnarray} 
A\bar{f}''+B\bar{f}'-\frac{Am}{\bar{f}}\bar{f}^{'2}-\bar{f}+\bar{f}^2=0, 
\label{gf61}
\end{eqnarray} 
with $\bar{f}=1-f, \;A=a^2(c_1^2n^2+c_2^2l^2),\; B=nl(c_1+c_2)$
and prime stands for differentiation with respect to $ \zeta.$ 
Because of the nonintegrable nature of Eq.~(\ref{gf61}), we 
look for  subcases by assuming one or more of the vector 
fields to be zero. In the following  
we report some of the nontrivial cases.

\subsubsection*{7.2(A)  \bfseries \emph{ Plane wave solutions}}

For $m=0$, making use of a similar procedure as in  the previous case
we can obtain the  plane wave solution. That is, $m=0$ leads to the ODE 
for $\bar{\zeta}=\zeta/\sqrt{A}$
\begin{eqnarray}
{\bar{f}}''+\frac{B}{\sqrt{A}}{\bar{f}}'-\bar{f}+\bar{f}^2=0.
\label{gf61a}
\end{eqnarray}
It is then straightforward to check that 
for $B/\sqrt{A}=5/\sqrt{6}$, Eq.~(\ref{gf61a}) satisfies the Painlev\'e property 
so that the solution 
to the original PDE is found to be
\begin{eqnarray}
u(x,y,t)=1-\Bigg[1+\frac{k}{\sqrt{6}}
\exp\Bigg(\frac{n(c_1+c_2)(ax-lt)-ac_2(ly-nx)}{\sqrt {6A}}\Bigg)\Bigg]^{-2}.
\label{gf61b}
\end{eqnarray}
Here we obtain a propagating plane wave and the pattern is 
plotted in (Figs.~9).
\begin{figure}[!ht]
\begin{center}
\epsfig{file=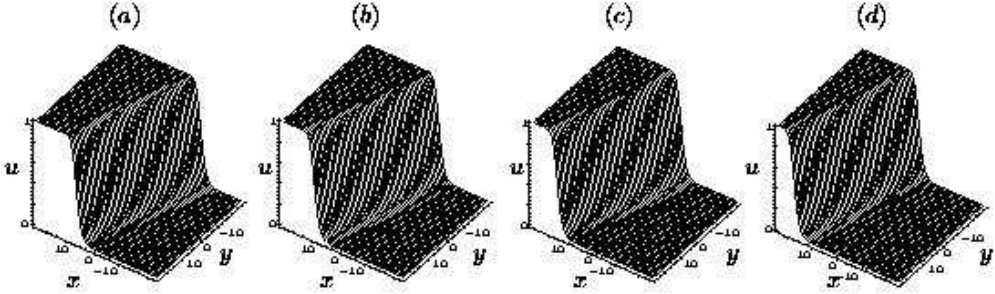,width=.9\linewidth}
\end{center}
\caption{ An exact propagating plane wave solution Eq.~(\ref{gf61b}) at 
different instants of time: (a) $t=0$, (b) $t=5$, (c) $t=10$, (d) $t=15$.}
\label{front2d}
\end{figure}

\subsubsection*{ 7.2(B) \bfseries \emph{Static and localized structures}}
Substituting $c_1=-c_2$ in (\ref{gf60}), we obtain the similarity 
transformations $\zeta=z_1,\;w=f(\zeta)$. This transforms 
Eq.~(\ref{gf59}) to an ODE 
\begin{eqnarray}
\bar{f}''-\displaystyle{\frac{m}{\bar{f}}}\bar{f}'{^2}-A\bar{f}
(1-\bar{f})=0,\;A=\frac{1}{l^2+n^2}.
\label{gf67a}
\end{eqnarray}
For Eq.~(\ref{gf67a}), in addition to the 
elliptic function solutions of the form tabulated in Table 1,
we obtain a particular planar solitary wave 
solution (Fig.~10)
\begin{eqnarray}
u=1-\frac{3-2m}{2-2m}\left[\mbox{sech}^2\left(I_2-
\frac{(nx-ly)}{2}\sqrt{\frac{A}{1-m}}\right)\right],\; m<1.
\label{gf68}
\end{eqnarray} 

We note the solutions given in (\ref{gf61b}) and (\ref{gf68}) are 
2-dimensional generalizations of (\ref{ls15}) and (\ref{ls21}) respectively.
\begin{figure}[!ht]
\begin{center}
\epsfig{file=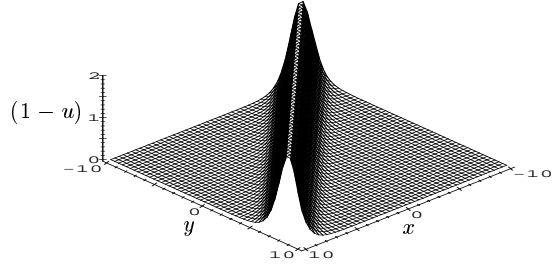,width=.5\linewidth}
\end{center}
\caption{Fig. 10.  A planar solitary wave solution for Eq.~(\ref{gf67a}) in the particular limit.}
\label{soln}
\end{figure}

Besides the above,  the choice $b=0,\;a=0,\;l=0$ reduces the 
(2+1) dimensional Fisher equation (\ref{gf1}) to that in (1+1) 
dimensions for $m\ne2$.
\begin{table}[!ht]
\caption{ Elliptic Function Solutions of Eq.~(\ref{ls17}) for  $m=1/2$.}
\begin{center}
\begin{tabular}{|c|c|c|c|}\hline
S.No. & Order of Roots & Function  $v(z)$ & modulus $k$ 
 \\ \hline
1. & $c_1 > c_2 > 0 \ge v $    & $v=c_1-\displaystyle{\frac{c_1}{{sn}^2\left(\displaystyle{\frac{\sqrt {c_1}}
{2}}(z-I_2),k\right)}} $ & $\displaystyle {k^2=\frac{c_1-c_2}{c_1}}$ \\

2. & $c_1 > c_2 > 0 > v $    & $v=\displaystyle{\frac
{c_2{sn}^2\left(\displaystyle{\frac{\sqrt {c_1}}
{2}}(z-I_2),k\right)}
{{sn}^2\left(\displaystyle{\frac{\sqrt {c_1}}
{2}}(z-I_2),k\right)-1}} $ & $\displaystyle {k^2=\frac{c_1-c_2}{c_1}}$ \\

3. & $c_1 \ge  v > c_2 > 0 $ & $v= \displaystyle{\frac{c_1c_2}{c_1-(c_1-c_2)sn^2
\left[\displaystyle{\frac{\sqrt {c_1}}{2}}(z-I_2),k 
\right]}}$ & $ k^2=\displaystyle{\frac{c_1-c_2}{c_1}}$\\

4.  & $c_1 >  v\ge c_2 > 0 $ & $v= c_1-(c_1-c_2)sn^2
\left[\displaystyle{\frac{\sqrt {c_1}}{2}}(z-I_2),k 
\right]$ & $ k^2=\displaystyle{\frac{c_1-c_2}{c_1}}$ \\ \hline
\end{tabular}
\label{table1}
\end{center}
\end{table}

\section{Nonclassical Reductions}
\label{s8}
Generally, for many PDEs there exist symmetry reductions that are not obtained 
by using the classical Lie group method. As a 
consequence, there have been several generalizations of the 
classical Lie group method for symmetry reductions. 
Bluman and Cole [1969], then 
proposed the
so-called \emph{non-classical method of group-invariant solutions}
in the study of symmetry reductions of a linear heat equation.
An algorithm  for calculating the determining equations associated
with the nonclassical method was presented by Clarkson and
Mansfield [1993]. This procedure has been applied to
several nonlinear systems and in some cases, such as the
Boussinesq equation, the Burgers' equation and the FitzHugh-Nagumo
equation, new similarity reductions  not obtainable through classical
symmetries have been found [Levi \& Winternitz, 1989; Arrigo \emph{et al.} 1993;
 Nucci \& Clarkson, 1992]
(see also Mansfield \emph{et al.} [1998] and references
therein). In the classical method we consider the infinitesimal 
generator, associated with the one-parameter Lie group of 
transformations,
\begin{eqnarray} 
V=\xi(t,x,u)\partial_x + \tau(t,x,u)\partial_t +
\phi(t,x,u)\partial_u 
\label{ncr1}
\end{eqnarray} 
which leaves the system (\ref{pain2}) invariant.
But in the nonclassical method, one 
 requires  the given equation 
(\ref{pain2})  and the  surface condition 
\begin{eqnarray}
\xi(t,x,u)u_x+\tau(t,x,u)u_t-\phi(t,x,u)=0 
\label{ncr2}
\end{eqnarray} 
together to be invariant
under the transformation with the infinitesimal 
generator 
(\ref{ncr1}). In this case one may obtain a larger
 set of solutions than that for the 
classical method. Moreover, significant progress has been made 
in the study of nonclassical symmetries for
nonlinear PDEs of diffusive type. To mention a few, Gandarias and Bruz\'on, [1999]
 have obtained several solutions for a family of 
Cahn-Hilliard  equations that are not invariant under
any Lie group admitted by the equation. In  a similar fashion,
separable new solutions that are not obtainable via 
classical method  
have been reported
for a mathematical model [Gandarias, 2001] of fast diffusion 
through nonclassical method. This prompts us to look for 
nonclassical symmetries associated with the GF equation (\ref{gfe2}).

\subsection{\emph{The (1+1) dimensional Fisher  equation}}
There are two types  of nonclassical symmetries: those where the
infinitesimal $\tau$ is non-zero and those where it is zero.  In
the first case, we can assume without loss of generality that
$\tau=1$, while in the second case we can assume  $\xi\ne0$ and $\tau=0$. In 
the following we investigate them separately.

\subsubsection*{8.1(A) \bf \emph{The case ${\bf \tau\ne0}$}}

We set $ \tau = 1 $ in the invariant surface condition without loss
of generality and use it together with its differential
consequences to eliminate $ u_t$ and so Eq.~(\ref{pain2}) takes the form 
\begin{eqnarray}
u_{xx}=\phi-\xi u_x-\frac {m}{1-u}u_x^2-u+u^2. 
\label{ncr3}
\end{eqnarray} 
Applying  the
classical Lie algorithm to (\ref{ncr3}) and eliminating  the
highest derivative involving the variable $ x $,  the
coefficients of the linearly independent expressions in the
remaining derivatives are set equal to zero. From the
resultant set of determining equations for $ \xi, \tau,$ and $
\phi $, which are in general nonlinear, we get 
\begin{subequations}
\begin{eqnarray} 
&&\tau  =  1,
\quad \xi  =   b, \quad \phi  = d(t,x)(1-u)^2, \quad m=2,
\label{ncr4a}\\
&&\tau = 1,\quad \xi = b,\quad \phi = 0, \quad m\ne2, \label{ncr4b}
\end{eqnarray}
\label{ncr4}
\end{subequations}
where $b$ is an arbitrary constant.
In Eq.~(\ref{ncr4a}), $d(t,x)$ satisfies $d_t-d_{xx}-d=0$. 

The corresponding similarity variables  for the $ m = 2 $ case are
\begin{eqnarray} 
z=x-bt,\quad  u=1- \displaystyle { \frac {1}{w(z)+\int d(t,x)dt}, }
\label{ncr5}
\end{eqnarray} 
which are similar to the ones obtained by the classical method
earlier in Sec.\ref{s4}. Thus we are lead to the similarity reduced differential equation
\begin{eqnarray} 
\displaystyle {\bar {w}''+b\bar {w}' + \bar {w}=0},\; \bar {w}= w - 1,\; '=d/dz,
\label{ncr6}
\end{eqnarray} 
which is similar to (\ref{ls10}) and the results then follows as before.
Similar is the case with $m\ne2$.  Thus both \emph{classical and nonclassical 
reductions lead to the same solutions} for the present system. 

\subsubsection*{8.1(B) \bf \emph{The case ${\bf\tau=0},\;{\bf\xi\ne0 }$}}

Let $ \tau = 0 $ and $ \xi \ne 0 $.  We then set $\xi = 1$, without loss of 
generality. The
invariance surface condition simplifies the Fisher equation to 
\begin{eqnarray}
u_t-\phi_x-\phi\phi_u-\frac{m}{1-u}\phi^2-u+u^2=0.
\label{ncr7}
\end{eqnarray} 
Applying
the classical Lie algorithm to the above equation we end up with a
more complicated nonlinear PDE
\begin{eqnarray}
\phi_t-\phi_{xx}-2\phi\phi_{xu}-\phi^2\phi_{uu}+u(1-u)\phi_u+(2u-1)\phi 
\nonumber \\ 
-\frac{2m}{1-u}\phi\phi_x -\frac{m}{1-u}\phi^2\phi_u-
\frac{m}{(1-u)^2}\phi^3&=0.
\label{ncr8}
\end{eqnarray}
This system is considerably more complex than the system 
(\ref{pain2}) and hence cannot be solved in general. 
We have made different ans\"atze for $\phi(t,x,u)$ and found that
all of them lead to only the group invariant solutions found by the classical algorithm.

However, for the $m=0$ case, by assuming $\phi$ to be independent of $x$ and $t$ we get $\phi = u\sqrt{\displaystyle{\frac{2u}{3}-1}},\; \tau=0,\; \xi=1$  which is different from that obtained through the classical method.  However this choice leads to a singular solution of the form 
\begin{eqnarray}
u=\frac{3}{2}\sec^2\left(\frac{x+c}{2}\right).
\end{eqnarray}

\subsection{\bfseries \emph{The (2+1) dimensional GF equation}}
We extend the  nonclassical
symmetry analysis to the (2+1) dimensional Fisher equation (\ref{gf1}) and
in this analysis there are three cases to consider, namely, (i) $ \tau\ne0
$, (ii) $ \tau = 0, \;\; \eta\ne0$, and
(iii) $\tau = 0, \;\; \eta = 0, \;\; \xi\ne0$.  

\subsubsection*{ 8.2(A) \bf \emph{The case $ {\bf\tau \ne 0} $ }}

We use the invariance surface condition 
\begin{eqnarray} 
u_t=\phi-\xi u_x-\eta u_y 
\label{ncr9}
\end{eqnarray} 
with its differential consequences to eliminate $ u_t $.
Applying the classical Lie algorithm to the Fisher equation
\begin{eqnarray}
\phi -\xi u_x -\eta u_y - u_{xx} -u_{yy} - \frac{m}{1-u}(u_x^2+u_y^2)
-u+u^2 = 0
\label{ncr10}
\end{eqnarray} 
and then equating coefficients of various powers of
derivatives of the dependent variable $ u $ we have a set of
determining equations. Solving them we get finally 
\begin{subequations}
\label{ncr11}
\begin{eqnarray} 
&&\tau=1,
\quad \xi=a_3y+a_4, \quad \eta = -a_3x+c_4,\quad \phi = d(t,x,y)
(1-u)^2,\quad m=2, \label{ncr11a} \\
&&\tau=1, \quad
\xi=a_3y+a_4, \quad \eta = -a_3x+c_4, \quad \phi = 0,\quad m\ne2. \label{ncr11b}
\end{eqnarray} 
\end{subequations}
Here $ d(t,x,y) $ satisfies the equation $d_t-d_{xx}-d_{yy}-d=0$ and 
$ a_3, a_4, c_4 $ are arbitrary constants. Comparing
this set of symmetries with that obtained by the classical method
we find that  they are similar to the latter case. So the
similarity reductions will reduce the PDE to the same ODE as that
obtained by the classical method. 

\subsubsection*{8.2(B) \bf \emph{The case $ {\bf\tau = 0} $, $ {\bf\eta \ne 0} $}}

Here we set $ \eta = 1 $, without loss of generality, and analogous to
the above procedure we use the above choice to eliminate all
$ y $-derivatives. Then applying the classical Lie algorithm to the
resulting equation we obtain a more complicated nonlinear PDE. 
Solving the resultant system of determining equations we obtain the 
following symmetries
\begin{subequations}
\label{ncr13}
\begin{eqnarray}
(i)  \quad && \tau=0,\quad\xi=c_1,\qquad \;\; \eta=1,\quad\phi=0,\label{ncr13a}\\
(ii) \quad &&\tau=0,\quad \xi=\pm\frac{1}{\sqrt{2}},\quad \eta=1,
\quad\phi=0,\;\label{ncr13b}
\end{eqnarray}
\end{subequations}
for all values of $m$. The similarity reduction variables $z_1=c_1y-x, 
\; z_2=t,\;u=w(z_1,z_2)$ associated with (\ref{ncr13a}) 
reduces the GF equation in 2-spatial dimensions to that in 
1-spatial dimension. Similarly, the similarity transformations 
$z_1=\pm\sqrt{2}y-x, \; z_2=t,\;u=w(z_1,z_2)$ for (\ref{ncr13b}) again reduces 
the original PDE  to that in 1-dimension and hence the 
results follow as in the case of the classical method.

\subsubsection*{ 8.2(C) \bf \emph{The case ${\bf\tau = 0}$, $ {\bf \eta = 0} $, $ 
{\bf\xi \ne0} $}}

In this case we set $ \xi = 1 $, without loss of generality, and
 as
above we make use of the invariance condition $ u_x = \phi $ to
eliminate all $ x $-derivatives. Then we apply the classical Lie
algorithm to the resulting equation and we get a nonlinear PDE of the form
\begin{eqnarray}
-\phi_t+\phi_{xx}+\phi_{yy}+\phi_{uu}(\phi^2+u_y^2)
+(1-2u+2\phi_{ux})\phi+2\phi_{uy}u_y
-u(1-u)\phi_u\nonumber\\
+\frac{m}{(1-u)}\Big [\frac{\phi}{(1-u)}(u_y^2+\phi^2)
+\phi_u(u_y^2+\phi^2)
+2\phi_yu_y+2\phi\phi_x\Big]&=0.
\label{ncr14}
\end{eqnarray}
Solving the system of determining equations we obtain $ \phi=0 $ 
$\forall \;m$. Thus the (2+1)-dimensional GF type equation,
under the similarity transformation
$z_1=y, \; z_2=t,\;u=w(z_1,z_2)$ reduces to that in 
(1+1)-dimensions and hence the  results as in the case 
of the classical Lie algorithm case follow.

\section{Generalized Conditional Symmetry Reductions}
\label{s9}
Recently Fokas  and  Liu [1994a,b]  proposed the notion of Generalized 
Conditional
Symmetry (GCS) and applied it to construct some physically interesting
exact solutions of certain nonlinear nonintegrable PDEs. Such exact 
solutions are of primary importance because they identify 
certain  interesting and  novel physical phenomena and 
moreover such solutions may be hard to identify or may not be quite 
transparent from the  numerical solution of a nonlinear PDE. In this method
the PDE is reduced  to an ODE in terms of certain types of generalized
conditional symmetries.  Indeed the GCS can be considered as 
 a natural generalization of
the nonclassical method just as  the generalized symmetry
method [Olver, 1986] is considered as a generalization of the Lie-point 
symmetry method.
Particularly, for the reaction diffusion type equations, several new
separable solutions that cannot be obtained via the non-classical and
non-local symmetry methods have been derived through this method 
[Qu, 1997; Qu, 1999a,b;  Chou \& Ku,  1999; Qu \emph{et al.}, 2000].
Further, it is recently shown that the GCS method is the most efficient
tool for solving the problem of dimensional reduction of initial value
problems for evolutionary type PDEs in a purely algebraic way 
[Zhadanov \& Andreitsev, 2000].  
Motivated by these facts we investigate the existence of generalized 
conditional symmetries, if any, for the GF equation.

\subsection{\bf \emph {The method}}
Let $K(t,u)$ denotes a function that depends on a differentiable
manner of $t, u, u_x, u_{xx}, \ldots$. The function $
\sigma(t,x,u) $ is a generalized symmetry of the evolution
equation $u_t=K(t,u)$ iff 
\begin{eqnarray} 
\frac{\partial
\sigma}{\partial t}+[K,\sigma]=0, 
\label{gcs1}
\end{eqnarray} 
when $ u_t = K(t,u) $, where
$ [K,\sigma] = \sigma'K - K'\sigma, $ and the prime denotes the
Gateaux derivative.   The function $ \sigma(t,x,u) $ is a GCS of
equation (\ref{pain2}) if there exists a function $ A $ such that 
\begin{eqnarray}
\frac{\partial \sigma}{\partial t}+[K,\sigma]=A(t,x,u,\sigma),
\quad A(t,x,u,0)=0, 
\label{gcs2}
\end{eqnarray} 
where $ K(t,u) $, and $ \sigma(t,x,u) $
are  differentiable functions  of $ t, x,u, u_x, u_{xx},\ldots$
and $ A(t,x,u,\sigma) $ is a differentiable function of $ t, x, u,
u_x,u_{xx}$,..., and $ \sigma, \sigma_x, \sigma_{xx},\ldots$ It is
obvious from (\ref{gcs2}) that Eq.~(\ref{gcs1}) admits a GCS $ \sigma $ iff 
\begin{eqnarray}
\sigma'K\mid_{\sigma=0} = 0, 
\label{gcs3}
\end{eqnarray}
provided $ \sigma $ is
explicitly independent of time $ t $.
\subsection{\bf \emph{The (1+1)- dimensional GF equation}}
Let the Fisher equation admits a GCS of the form 
\begin{eqnarray} 
\sigma =u_{xx}+H(u)u_x^2+F(u)u_x+G(u), 
\label{gcs4}
\end{eqnarray} 
where $H(u),F(u)$ and $G(u)$
are arbitrary functions $u$.  Substituting (\ref{gcs4}) into (\ref{gcs3}) one gets
 \begin{eqnarray}
\left[H_{uu}-2H(2H_u-H^2)-\frac{m}{(1-u)^2}\left(\frac{2}{1-u}-3
H\right)
+\frac{m}{1-u}\left(H_u-2H^2\right)\right]
u_x^4\nonumber\\
+\Bigg[F_{uu}-2HF_u-4F(H_u-H^2)+\frac{m} {1-u}(F_u+4FH)+
 \frac{4mF}{(1-u)^2}\Bigg]u_x^3 
 \nonumber\\  -\Bigg[G_{uu}-2F(F_u-HF)
-4G(H_u-H^2)+(2u-1)H-u(1-u)H_u\Bigg.\nonumber\\\Bigg.+\frac{m}{1-u}\left(4HG-2F^2+G_u\right) 
+\frac{5mG}{(1-u)^2}\Bigg]u_x^2- \Bigg[
\left(u(1-u)+2G\right)F_u
 \Bigg.\nonumber \\\Bigg.+4FG\left(\frac{m}{1-u}-H\right)\Bigg]u_x 
- u(1-u)G_u +(1-2u)G
\nonumber\\+2G^2(H-\frac{m}{1-u})&=0. 
\label{gcs5}
\end{eqnarray} 
Equating coefficients of various powers of $ u_x $  to
zero and then solving the resulting overdetermined equations  we
get 
\begin{eqnarray}
H(u)= \displaystyle{\frac{m}{1-u}},\quad G(u)= 
u(1-u),\quad  F(u)= c_1,\quad \forall\; m,
\label{gcs6}
\end{eqnarray} 
where $ c_1 $ and $ c_2 $ are arbitrary
constants.
The corresponding GCS are 
\begin{eqnarray}
\sigma=u_{xx}+\frac{m}{1-u}u_x^2+c_1u_x+u-u^2.
\label{gcs7}
\end{eqnarray}
With the above some exact solutions can be constructed. The
ODE $\sigma = 0$ is solved to obtain $ u $ as a function 
of $x$ with integration constants dependent on time $t$ 
alone. Then substituting this solution into the governing 
equation, the time evolution of these time-independent 
constants are determined. The exact solutions are then 
obtained by solving these systems.

The GCS (\ref{gcs7}) leads to
\begin{eqnarray}
u''+\frac{m}{1-u}u'^2+c_1u'+u-u^2=0
\label{gcs8}
\end{eqnarray}
which cannot be integrated as such except for $m=2$.
Thus for $m=2$, we obtain
\begin{eqnarray} 
u=1-\left[1+A(t)e^{m_1x}+B(t)e^{m_2x}\right]^{-1},\quad  m_{1,2}=\frac{-c_1\pm\sqrt{c_1^2-4}}{2}.
\label{gcs9}
\end{eqnarray}
Substituting (\ref{gcs9}) into the Fisher equation (\ref{pain2}) we get 
\begin{subequations}
\label{gcs10}
\begin{eqnarray} 
&\displaystyle{u=1-\left(1+Ae^{\displaystyle[(m_1^2+1)t+m_1x]}+Be^{\displaystyle[(m_2^2+1)t+m_2x]}\right)^{-1} },\label{gcs10a} \\&\displaystyle{m_{1,2}=\frac {c_1 \pm \sqrt{c_1^2-4c_2}}
{2}}, &
\label{gcs10b}
\end{eqnarray} 
\end{subequations}
for $m=2$. 
We wish to mention that the solution (\ref{gcs10a}) can be recovered
from the linear heat equation Wang \emph{et al.} [1996] and the results are similar to that obtained through the classical algorithm. Further Eq.~(\ref{gcs10b}) is similar to Eq.~(\ref{ls12}) and hence we 
obtain the various  propagating structures already discussed through the classical method.

Eq.~(\ref{gcs8}) is of non-Painlev\'e type for all other values of m ($\ne2$)
and  is similar to (\ref{ls14}) and hence the previous results follow.
Thus GCS reductions lead to the same results as that obtained through
the classical method for the modified perturbed GF equation.

\section{Conclusions}
\label{con}
In this paper, we have investigated the integrability/symmetry 
properties of the generalized Fisher type nonlinear 
reaction-diffusion equation in both (1+1) and (2+1) dimensions.
The singularity structure analysis singles out the $m=2$ case 
as the only system parameter for which the GF equation is free 
from movable critical singular manifolds. Further a B\"acklund transformation for the integrable case is shown to linearize the GF equation to the linear heat equation. The symmetry analysis shows that the system
under consideration possesses interesting Lie point symmetries that lead to the infinite-dimensional Lie algebra for the integrable
case, thereby exhibiting various interesting patterns and dynamics, and  giving rise to a group theoretical interpretation for the system. In addition to the above, we have given exact and numerical solutions for specific choices of the nonintegrable case. Our studies 
on the GF equation reveals the existence of a large number of interesting wave patterns, static and localized structures. We have also carried out the nonclassical and generalized conditional symmetry reductions.

\section*{Acknowledgments}
This work forms a part of the National Board of Higher Mathematics,
Department of Atomic Energy, Government of India and the Department of
Science and Technology, Government of India research projects.

\section*{References}
\begin{description}
\item
 Ablowitz, M.J.  \&  Zeppetella, A. [1979] "Explicit solution of Fisher's equation for a special
 wave speed", \emph{Bull. Math. Biol.}, {\bf 41}, 835-840.

\item
Ablowitz,~M. J. \& Clarkson,~P. A. [1991]  
"\emph{Solitons: Nonlinear Evolution Equations and Inverse
Scattering}"  (Cambridge University Press, Cambridge).

\item
Arrigo, D.J.,  Broadbridge, P.  \&  Hill, J.M. [1993] "Nonclassical symmetry solutions and the
methods of Bluman-Cole and Clarkson and Kruskal", \emph{J. Math. Phys.}, {\bf 34}, 4692-4703.

\item
 Bindu, P.S.,   Senthilvelan, M. \&  Lakshmanan, M. [2001] "Singularity structures, symmetries and
 integrability aspects of generalized Fisher type nonlinear diffusion equation", 
 \emph{J.~Phys.~A:~Math.}\\\emph{~Gen.}, {\bf 34}, 
 L689-L696.
 
 \item
Bindu, P.S.  \& Lakshmanan, M. [2002] "{Symmetries and integrability properties of 
generalized Fisher 
type nonlinear diffusion equation}, \emph{Proceedings of the Institute of Mathematics of
 NAS, Ukraine}, Eds. Nikitin,
A.G., Boyko, V.M. \& Popovych, R.O., Part -1, {\bf 43}, 36-48.

\item
Bluman, G.W. \&  Cole, J.D. [1969] "The general similarity solutions of the heat equation", 
\emph{J. Math. Mech.}, {\bf 18},
1025-1042.

\item
Bluman, G.W. \&  Kumei, S. [1989] "\emph{Symmetries and Differential
Equations}" (Springer-Verlag, New York).

\item
 Brazhnik, P.K.  \&  Davydov, V.A. [1995] "Non-spiral autowave structures in unrestricted
 excitable media", \emph{Phys.~Lett.~A.}, {\bf 199}, 40-44.
 
 \item
 Brazhnik, P.K.,  Fan, S. \&   Tyson, J.J. [1996] "Nonspiral excitation waves beyond the eikonal
 approximation", 
\emph{Phys.~Rev.~E.}, {\bf 54}, 4338-4346.

\item
 Brazhnik, P.K.  \&  Tyson, J.J. [1999a] "Travelling waves and static structures in a
 two-dimensionsal exactly solvable reaction-diffusion system",
\emph{J.~Phys.~A:~Math.~Gen.}, {\bf 32},
8033-8044.

\item
Brazhnik,P.K.  \& Tyson, J.J. [1999b] "On travelling wave solutions of Fisher's equation in two
spatial dimensions",
\emph{SIAM.~J.~Appl.~Math.}, {\bf 60}, 371-391.

\item
 Calogero, F. [1991] "{Why are Certain Nonlinear PDEs Both Widely Applicable and 
 Integrable?}",  
 in \emph{What is Integrability?} Ed.~V.~E.~ Zakharov,  (Springer-Verlag, Berlin), p1-62.

\item
 Chou, K.S. \& Qu, C.Z. [1999] "Symmetry groups and separation of variables of a class of
 nonlinear diffusion-convection equations", \emph{J. Phys. A: Math. Gen.}, {\bf 32},
6271-6286.

\item
Clarkson, P.A. \&  Mansfield, E.L. [1993] "Symmetry reductions and exact solutions of a class of
nonlinear heat equations", \emph{Physica D}, {\bf 70},
250-288.

\item
Fokas, A.S. \&  Yortsos, Y.C. [1982]  "On the exactly solvable equation 
$S_t=[(\beta S+\gamma)^{-2}S_x]_x+\alpha(\beta S+\gamma)^{-2}S_x$ occuring in two-phase flow in
porous media", \emph{SIAM J. Appl. Math.},
{\bf 42}, 318-332.

\item
Fokas, A.S.  \& Liu, Q.M. [1994a]  "Nonlinear interaction of travelling waves of
nonintegrable equations", \emph{Phys. Rev. Lett.}, {\bf 72},  
3293-3296.

\item
Fokas, A.S.  \& Liu, Q.M. [1994b] "Generalized conditional symmetries and exact solutions
of nonintegrable equations", \emph{Theor. Math. Phys.}, {\bf 99}, 
263-277.

\item
Gandarias, M.L.  \& Bruz\'on, M.S. [1999] "Nonclassical symmetries for a family of
Cahn-Hilliard equations", \emph{Phys. Lett. A}, {\bf 263}, 331-337.

\item
Gandarias, M.L. [2001] "New symmetries for a model of fast diffusion", 
\emph{Phys. Lett. A}, {\bf 286}, 153-160.

\item
Grimson, M.J. \& Barker, G.C. [1994] "Continuum model for the spatiotemporal growth of bacterial
colonies",   \emph{Phys.~Rev.~E.}, {\bf 49}, 1680-1684.

\item
Head, A.  [1993] "LIE: a PC program for Lie analysis of differential equations", 
\emph{Comput. Phys. Commun.}, {\bf 77}, 241-248.

\item
Lakshmanan, M. \& Rajasekar, S. [2003] "\emph{Nonlinear Dynamics: Integrability, Chaos, and Patterns}",
(Springer-Verlag, Berlin).
\item
Levi, D. \& Winternitz, P. [1989] "Non-classical symmetry reduction: example of Boussinesq
equation",  \emph{J. Phys. A: Math. Gen.},
{\bf 22}, 2915-2924.

\item
 Mansfield, E.L., Reid,  G.J. \& Clarkson,  P.A. [1998] "Non-classical reductions of a (3+1) dimensional cubic
 nonlinear Schr\"odinger equation", 
\emph{Compt. Phys. Comm.}, {\bf 115}, 460-488.

\item
  Mathews, P.M. \& Lakshmanan, M.  [1974] "On a unique nonlinear oscillator", 
 \emph{Qt. Appl. Math.}, {\bf 32}, 215-218.

\item
 Murphy, G.~M.~[1969] "\emph{Ordinary Differential 
Equations and Their Solutions}" (Affiliated East-West Press, New 
Delhi).

\item 
Murray, J.D. [1989] "\emph{Mathematical Biology}"
 (Springer-Verlag, Berlin).

\item
Nucci, M.C. \&  Clarkson, P.A. [1992] "The nonclassical method is more general than the direct
method for symmetry reductions: An example of the FitzHugh-Nagumo equation", \emph{Phys. Lett. A}, 
{\bf 164}, 56-59.

\item
Olver, P.J. [1986] "\emph{Applications of Lie Groups to Differential
Equations}" (Springer-Verlag, New York).

\item
 Qu, C.Z. [1997] "Group classification and generalized conditional symmetry reduction of the
 nonlinear diffusion-convection equation with a nonlinear source", 
 \emph{Stud. Appl. Math.}, {\bf 99}, 107-136.

 \item
 Qu, C.Z. [1999a] "New generalized conditional symmetry reductions and exact solutions of
 the nonlinear diffusion-convection-reaction equations", 
 \emph{Commun. Theor. Phys.}, {\bf 31}, 581-588.
 
 \item
 Qu, C.Z. [1999b] "Reductions and exact solutions of some nonlinear partial differential
 equations under four types of generalized conditional symmetries", 
  \emph{J. Austral. Math. Soc. B},  {\bf 40}, 1-42.
  
  \item
 Qu, C.Z., Zhang, S. \& Liu, R.  [2000] "Separation of variables and exact solutions to quasilinear diffusion
 equations with nonlinear source", 
 \emph{Physica D}, {\bf 144}, 97-123.
 
 \item
Rosen, G. [1982] "Method for the exact solution of a nonlinear diffusion-convection equation",
 \emph{Phys. Rev. Lett.}, {\bf 49}, 1844-1847.

\item
 Sachdev, P.L. [1987]  "\emph{Nonlinear Diffusive Waves}"
(Cambridge University Press, Cambridge).

\item
 Schwendeman, D.W.  [1996] "A front dynamics approach to curvature-dependent flow", 
\emph{SIAM.~J.}\\\emph{~Appl.~Math.}, {\bf 56}, 1523-1538.

\item
 Scott, S.K. \& Showalter, K.  [1992] "Simple and complex propagating reaction-diffusion front", 
  \emph{J.~Chem.~Phys.}, {\bf 96}, 8702-8711.
  
  \item
Scott, A.C. [ 1999] "\emph{Nonlinear Science: Emergence and Dynamics
of Coherent Structures}"  (Oxford University Press, Oxford).

 \item
Showalter, K.  [1995] "Quadratic and cubic reaction-diffusion fronts", 
\emph{Nonlinear.~Sci.~Today}, {\bf 4}, 1-10.

\item
Walgraef, D. [1996] "\emph{Spatiotemporal Pattern Formation}" (Springer-Verlag, NewYork).

\item
 Wang, X.Y., Fan,  S. \& Kyu, T. [1996] "Complete and exact solutions of a class of nonlinear
 diffusion equations and problem of velocity selection",
\emph{Phys.~Rev.~E.}, {\bf 56}, R4931-R4934.

\item
 Weiss, J., Tabor, M.  \& Carnevale, G. [1983] "The Painlev\'e property for
  partial differential
 equations", \emph{J. Math. Phys.}, {\bf 24}, 522-526.

\item
 Whitham, G.B. [1974] "\emph{Linear and Nonlinear Waves}"
 (Wiley, NewYork).
 
 \item
 Zhadanov, R.Z. \& Yu Andreitsev, A.  [2000]  "Non-classical reductions of intial-value
 problems for a class of nonlinear evolution equations", \emph {J. Phys. A: Math. Gen.}, {\bf 33}, 5763-5781.

\end{description}

\end{document}